\documentclass[12pt]{article}

\usepackage{a4wide,epsfig,psfrag,cite}
\usepackage{amsmath,amssymb,scalefnt}
\usepackage{color}
\usepackage{slashed}
\usepackage{tikz}
\usetikzlibrary{shapes.geometric, arrows}
\usetikzlibrary{positioning}
\tikzstyle{rec} = [rectangle, minimum width=0.5cm, minimum height=0.5cm, text centered, draw=black]
\tikzstyle{ell} = [ellipse, minimum width=0.5cm, minimum height=0.5cm, text centered, draw=black]
\tikzstyle{arrow} = [thick,->,>=stealth]

\graphicspath{ {figs/} }

\definecolor{darkgreen}{cmyk}{1.0,0,1.0,0.61}

\definecolor{light-gray}{gray}{0.95}

\allowdisplaybreaks

\begin{document}

\title{\vskip-3cm{\baselineskip14pt
    \begin{flushleft}
      \normalsize TTP17-046
  \end{flushleft}}
  \vskip1.5cm
  Gauge and Yukawa coupling beta functions of
  two-Higgs-doublet models to three-loop order
}

\author{
  Florian Herren$^{(a)}$,
  Luminita Mihaila$^{(b)}$
  and
  Matthias Steinhauser$^{(a)}$
  \\[1em]
  {\small\it ${(a)}$ Institut f{\"u}r Theoretische Teilchenphysik}\\
  {\small\it Karlsruhe Institute of Technology (KIT)}\\
  {\small\it 76128 Karlsruhe, Germany}\\[1em]
  {\small\it ${(b)}$ Institut f{\"u}r Theoretische Physik}\\
  {\small\it Universit{\"a}t Heidelberg}\\
  {\small\it 69120 Heidelberg, Germany}
}

\date{}

\maketitle

\begin{abstract}
  We compute the beta functions for the three gauge couplings and the Yukawa
  matrices of a general two-Higgs-doublet model in the
  modified minimal subtraction scheme to three loops. The calculations are
  performed using Lorenz gauge in the unbroken phase.
  We discuss in detail the occurence of poles in anomalous dimensions
  and propose practical prescriptions to avoid them. We provide explicit
  results for the often used $\mathbb{Z}_2$-symmetric versions of the two-Higgs-doublet
  model of type I, II, X and Y.
  Furthermore, we provide the first independent cross-check of the
  three-loop Yukawa coupling beta functions of the Standard Model.

  \medskip

\end{abstract}

\thispagestyle{empty}


\newpage


\section{Introduction}

An appealing renormalization scheme for the couplings of the Standard Model of
particle physics (SM) and of its extensions is the minimal modified
subtraction ($\overline{\text{MS}}$) scheme. As a consequence, the numerical
values of the couplings depend on the renormalization scale $\mu$, which in general is of the
same order as the energy scale of the considered process. The values of the
couplings at different scales are related by so-called beta functions which in
perturbation theory are given as power series in all couplings of the theory.

In the SM there are three gauge couplings ($g_1, g_2, g_s$), the quartic Higgs
boson coupling $\lambda$, and a Yukawa coupling for each massive fermion,
where often only the third generation couplings, $y_t$, $y_b$, and $y_\tau$ are
considered as non-zero. For all couplings the three-loop beta functions have
been completed recently: the gauge coupling beta functions have been computed
in Refs.~\cite{Mihaila:2012fm,Mihaila:2012pz,Bednyakov:2012rb}, the ones for
the Yukawa couplings in
Refs.~\cite{Bednyakov:2012en,Bednyakov:2013cpa,Bednyakov:2014pia} and
$\lambda$ has been considered in~\cite{Chetyrkin:2013wya,Bednyakov:2013eba}.
Leading terms to the four-loop QCD beta function and the Higgs self coupling
involving the top Yukawa coupling and $\alpha_s$ have been computed in
Refs.\cite{Martin:2015eia,Bednyakov:2015ooa,Zoller:2015tha,Chetyrkin:2016ruf} 
and within QCD the beta function is even known to five
loops~\cite{Baikov:2016tgj,Herzog:2017ohr,Luthe:2017ttg}.

There are a number of two-loop results which can be immediately adapted to a
large class of non-supersymmetric beyond-the-SM theories.
In particular, two-loop results for
gauge~\cite{Machacek:1983tz}, Yukawa~\cite{Machacek:1983fi} and scalar
self couplings~\cite{Machacek:1984zw} are known since
middle of the eighties. Furthermore, the three-loop gauge coupling beta
function for a simple gauge group has been calculated \cite{Pickering:2001aq}.
In this work we consider the so-called two-Higgs-doublet model
(2HDM) and compute the gauge and Yukawa coupling beta functions
to three-loop order.

2HDMs, where the SM Higgs sector is extended by a second $SU(2)$ Higgs doublet,
are attractive extensions of the SM. Although simple and probably not realized
in nature in its minimal version, 2HDMs nevertheless constitute
prototype-extensions of the SM which can be used to study several features of
beyond-SM theories. In particular, for a certain choice of parameters it
implements the Higgs sector of the Minimal Supersymmetric Standard
Model. Further motivation and several phenomenological applications can be
found in the review~\cite{Branco:2011iw}.

The most general 2HDM has many parameters and furthermore
several unwanted features like flavour-changing neutral currents (FCNCs) at tree
level. Thus, often additional symmetries are imposed.
For example, if CP conservation in the Higgs sector is assumed
one has five physical scalar degrees of freedom which correspond
to two scalar, one pseudo scalar and a charged Higgs boson.
In these models, both Higgs doublets acquire vacuum
expectation values $v_{1}$ and $v_{2}$ such that $v = \sqrt{v_1^2 + v_2^2} \simeq
246$~GeV determines the $W^\pm$ and $Z$ boson masses in the same way as in
the SM.  The ratio $v_2/v_1$ is denoted by $\tan\beta$.

The scope of the present work is twofold: First, we provide the first
independent cross check of the three-loop Yukawa coupling beta functions in
the SM. In this context it is particularly important to carefully investigate
the scheme used for $\gamma_5$ in $D\not=4$ dimensions.  Note that for the
gauge couplings it is possible to choose Green's functions without external
fermions. For Yukawa couplings this is not possible anymore.
As a second aim, we extend both the gauge and Yukawa beta functions to a
general 2HDM.  There is no change in the underlying integrals, which have to
be evaluated, however, there are conceptional challenges in connection to the
wave function renormalization of the scalar fields.

The remainder of the paper is organized as follows: In the next section we
introduce the 2HDM which serves to fix the notation. Section~\ref{sec:tech} is
devoted to technical details. In particular, we introduce the renormalization
constants for the parameters and fields and define the beta functions and
anomalous dimensions which we want to compute. The main focus of Section~\ref{sec:ren}
relies on the proper definition of the renormalization constants such that the anomalous
dimensions are finite. We investigate this problem in detail and propose practical solutions.
A detailed discussion of the
computation of the gauge and Yukawa coupling beta functions is provided in
Sections~\ref{sec:gau} and~\ref{sec:yuk}, respectively.  In these Sections we
also explain how one can arrive at special versions of the 2HDM and the SM
results.  Furthermore, we compare the Yukawa beta functions to
Ref.~\cite{Bednyakov:2012en}.  The findings of this paper are summarized in
Section~\ref{sec:sum}.


\section{\label{sec:2HDM}Two-Higgs-doublet model}

An extensive discussion of a general 2HDM model can be found in
Ref.~\cite{Branco:2011iw}.  For convenience we repeat in the following the
features which are important for our calculation.

The additional Higgs doublet leads to an enlarged Yukawa sector
which can be written as
\begin{align}
  \mathcal{L}_Y = - \left(\sum_{i = a}^2 
  \bar{Q}_{L}\tilde{\Phi}_a Y^u_{a} u_{R} 
  + \bar{Q}_{L} \Phi_a Y^d_{a} d_{R} 
  + \bar{L}_{L} \Phi_a Y^l_{a} l_{R} 
  + \mathrm{h.c.}\right)\,.
  \label{eq::yuk}
\end{align}
The sum runs over the two doublets and ``h.c.'' refers
to the hermitian conjugate part. $Y^u_{a}$,$Y^d_{a}$ and $Y^l_{a}$ are generic
$3\times3$ complex matrices containing the Yukawa couplings and $Q_L$, $L_L$,
$u_R$, $d_R$ and $l_R$ represent left- and right-handed quark and lepton
fields.  $\tilde{\Phi} = i\tau_2 \Phi_j^*$ is the charge conjugated doublet
with $\tau_2$ being the second Pauli matrix.

The 2HDM has furthermore a more involved scalar potential which in its general
form is given by~\cite{Wu:1994ja}
\begin{eqnarray}
  V\left(\Phi_1,\Phi_2\right) &=& m_{11}^2\Phi_1^\dagger \Phi_1 +
  m_{22}^2\Phi_2^\dagger \Phi_2 - \left(m_{12}^2\Phi_1^\dagger \Phi_2 +
    \mathrm{h.c.}\right)
  +\frac{1}{2}\lambda_1\left(\Phi_1^\dagger\Phi_1\right)^2
  \nonumber\\ 
  &&\mbox{}
  +\frac{1}{2}\lambda_2\left(\Phi_2^\dagger \Phi_2\right)^2
  + \lambda_3\left(\Phi_1^\dagger \Phi_1\right)\left(\Phi_2^\dagger
    \Phi_2\right) + \lambda_4\left(\Phi_1^\dagger
    \Phi_2\right)\left(\Phi_2^\dagger \Phi_1\right)
  \nonumber\\
  &&\mbox{}+\left[\frac{1}{2}\lambda_5\left(\Phi_1^\dagger
      \Phi_2\right)^2+\lambda_6\left(\Phi_1^\dagger
      \Phi_2\right)\left(\Phi_1^\dagger
      \Phi_1\right)+\lambda_7\left(\Phi_1^\dagger
      \Phi_2\right)\left(\Phi_2^\dagger \Phi_2\right)+
    \mathrm{h.c.}\right]\,.
  \nonumber\\*
  \label{eq::V}
\end{eqnarray}
The parameters $m_{11}^2$, $m_{22}^2$ and $\lambda_1$, \ldots, $\lambda_4$ are
real whereas in general $m_{12}^2$, $\lambda_5$, $\lambda_6$ and $\lambda_7$
are complex. This leads to fourteen degrees of freedom, eleven of which are physical
as can be seen by an appropriate basis choice for $\Phi_1$ and
$\Phi_2$~\cite{Branco:2011iw}.

As in all multi-Higgs-doublet models, the Lagrange densities given in
Eqs.~\eqref{eq::yuk} and \eqref{eq::V} contain FCNCs. For example, the
up-type Yukawa matrices $Y^u_{1}$ and $Y^u_{2}$ will not be in general
simultaneously diagonalizable and thus neutral Higgs scalars $\phi$
will mediate FCNCs of the form $\bar{u} u^\prime \phi$ already at the tree level, where $u\ne
u^\prime$ are two different up-type quarks.  To avoid FCNCs at tree
level~\cite{Paschos:1976ay,Glashow:1976nt} it is necessary that all
fermions with the same quantum numbers couple to one and the same
Higgs multiplet. This condition can be satisfied if all quarks couple
to just one of the Higgs doublets or the right-handed up- and
down-type quarks couple to different Higgs doublets. Depending on
whether the right-handed leptons couple to the Higgs doublets in the
same manner as the right-handed down-type quarks, or in the opposite
way, further two possibilities can be identified. The resulting four models
are summarized in Tab.~\ref{tab::types}.
They can be realized by imposing a $\mathbb{Z}_2$ symmetry to the general model.
In fact, the type I 2HDM can be obtained by enforcing an additional $\mathbb{Z}_2$
symmetry under which the theory has to be invariant namely $\Phi_1 \to
-\Phi_1$ and $\Phi_2 \to \Phi_2$.  The type II 2HDM can be derived via
the symmetries $\Phi_1 \to -\Phi_1$, $\Phi_2 \to \Phi_2$, $d_R\to
-d_R$ and $l_R\to -l_R$. The additional discrete symmetries required
for the other two models can be derived similarly. Note that the
$\mathbb{Z}_2$ symmetries require that $m_{12}=\lambda_6=\lambda_7=0$.

In a generic quark basis as given in Eq.~\eqref{eq::yuk} the condition
for non-existence of FCNCs in the up-type (down-type) quark sector is that
the Yukawa matrices $Y^u_{1}$ and $Y^u_{2}$ ($Y^d_{1}$ and $Y^d_{2}$)
commute~\cite{Ecker:1987qp}. If one of the two Yukawa matrices is
zero, as it is actually the case for the four models shown in
Tab.~\ref{tab::types}, this condition is trivially fulfilled. 

The most general Lagrange densities in Eqs.~\eqref{eq::yuk}
and \eqref{eq::V} contain several fields with the same quantum
numbers that can mix. Therefore, one can rewrite the Lagrangian in
terms of the new fields obtained from the original ones by simple
basis transformations. In the following we will refer to these
transformations as flavour transformations for both fermions and
scalars. Obviously, the physical observables do not depend on such
redefinitions. They can depend only on quantities that are invariant
under arbitrary unitary flavour transformations. Ideally, one would be
able to express the fundamental Lagrangian parameters in terms of
these invariants. However, some of the Lagrangian parameters in
Eq.~\eqref{eq::yuk}, that do not take into account flavour
symmetries are not physical. That is, there are Lagrangian parameters
that can be expressed as linear combinations of others.  This
also means that there is a basis where the unphysical parameters are
identically zero, i.e. one can rotate them away via flavour
transformations. In other words, any coupling or mixing angle can be
expressed in terms of so-called flavour invariants. This statement has been
explicitly proven for the Yukawa sector of the
SM~\cite{Feldmann:2015nia} and for the scalar sector of the 2HDM, for
example, in Refs.~\cite{Botella:1994cs,Davidson:2005cw}.\footnote{For
  more details see Ref.~\cite{Branco:2011iw} and references therein.}
In this paper, we (re)confirm the findings of~\cite{Feldmann:2015nia,Botella:1994cs,Davidson:2005cw}
explicitly for the Yukawa sector of
the SM and for $\mathbb{Z}_2$-symmetric 2HDMs through three loops.

\begin{table}[t]
  \begin{center}
    \begin{tabular}{c|ccc}
      Type & $u_R$ & $d_R$ & $l_R$ \\
      \hline
      I & $\Phi_2$ & $\Phi_2$ & $\Phi_2$ \\
      II & $\Phi_2$ & $\Phi_1$ & $\Phi_1$ \\
      X & $\Phi_2$ & $\Phi_2$ & $\Phi_1$ \\
      Y & $\Phi_2$ & $\Phi_1$ & $\Phi_2$ \\
    \end{tabular}
    \caption{\label{tab::types}Four $\mathbb{Z}_2$-symmetric 2HDMs.
      The table shows which right-handed fermion field couples to which
      doublet.}
  \end{center}
\end{table}

The flavour transformations for fermion and  scalar fields of the in Lagrangian
Eq.~(\ref{eq::yuk}) can be summarized as follows
\begin{eqnarray}
  Q'_{L,I} &=& U_{Q,IK}Q_{L,K}~,\nonumber\\
  u'_{R,i} &=& U_{u,ik}u_{R,k}~,\nonumber\\
  d'_{R,m} &=& U_{d,mp}d_{R,p}~,\nonumber\\
  \Phi'_{a} &=& U_{\Phi,ac}\Phi_{c}~,
  \label{eq::trafo}
\end{eqnarray}
where $U_{Q}$, $U_{u}$ and $U_{d}$ are unitary $3\times 3$ matrices
and $U_{\Phi}$ is a unitary $2\times 2$ matrix.  Under these unitary
basis transformations, the gauge and kinetic terms are unchanged
and $\mathcal{L}_Y$ in Eq.~(\ref{eq::yuk}) is invariant if the Yukawa
matrices transform as
\begin{eqnarray}
  Y^{d \prime}_{a,Im} &=&
  U_{Q,IK}Y^d_{b,Kp}U^\dagger_{d,pm}U_{\Phi,ba}^\dagger~\,\nonumber\\ Y^{u
    \prime}_{a,Ij} &=&
  U_{Q,IK}Y^u_{b,Kl}U^\dagger_{u,lj}U_{\Phi,ba}^{\rm T}.
  \label{eq::trafo2}
\end{eqnarray}
In a similar manner, one can derive the transformation properties of
the parameters in the potential under redefinitions of the scalar
fields~\cite{Branco:2011iw}. One introduces the rank two and four
tensors, $K_{ab}$ and $\lambda_{ab,cd}$, so that
\begin{eqnarray}
  V\left(\Phi_1,\Phi_2\right) &=& K_{ab} \Phi_a^\dagger \Phi_b + \frac{1}{2}
  \lambda_{ab,cd} (\Phi_a^\dagger \Phi_b)(\Phi_c^\dagger \Phi_d)\,,
\end{eqnarray}
with
\begin{eqnarray}
  K_{ab}&=&K_{ba}^\ast\,,\quad \lambda_{ab,cd}
  =\lambda_{cd,ab}\,,\quad \lambda_{ab,cd} =\lambda_{ba,dc}^\ast\,. 
\end{eqnarray}
One can match with the standard notation given in Eq.~(\ref{eq::V}) and obtain
the following relations
\begin{align}
  &&K_{11} =  m_{11}^2\,,
  &&&K_{12} = - m_{12}^2\,,
  &&K_{21} = - (m_{12}^2)^\ast\,,
  \nonumber\\
  &&\lambda_{11,11}=\lambda_1\,,
  &&&\lambda_{22,22}=\lambda_2\,,
  &&\lambda_{11,22}=\lambda_{22,11}=\lambda_3\,,
  \nonumber\\
  &&\lambda_{12,21}=\lambda_4\,,
  &&&\lambda_{12,12}=\lambda_5\,,
  \nonumber\\
  &&\lambda_{11,12}=\lambda_6\,,
  &&&\lambda_{22,12}=\lambda_7\,.
 \label{eq::lamtransf}
\end{align}
The two tensors transform under the basis change given in Eq.~(\ref{eq::trafo}) as
\begin{eqnarray}
  K_{ab}^\prime &=& U_{\Phi,a\alpha} K_{\alpha\beta} U_{\Phi,\beta
    b}^\dagger\,,\nonumber\\ 
  \lambda_{ab,cd}^\prime&=& U_{\Phi,a\alpha}  U_{\Phi,c \rho}
  \lambda_{\alpha\beta,\rho\sigma} U_{\Phi,\beta b}^\dagger U_{\Phi,\sigma
    d}^\dagger\,.
 \label{eq::trafo3}
\end{eqnarray}
Since the calculation of the $\overline{\text{MS}}$ renormalization constants can be performed
in the unbroken phase the dimensionful parameters
$m_{ij}$ are irrelevant and thus for our calculation of the beta
functions only the second transformation in Eq.~\eqref{eq::lamtransf} will be of interest.

Within the SM the physical Yukawa couplings are defined via the
diagonalization of the hermitian matrices 
\begin{eqnarray}
M_{u,1} = Y^u_1Y^{u \dagger}_1 =  U_{uL,1} D_{u,1}^2  U_{uL,1}^\dagger \,,\quad
Y^{u \dagger}_1Y^u_1 =  W_{uR,1} D_{u,1}^2  W_{uR,1}^\dagger \,,
\label{eq::diag}
\end{eqnarray}
where $U_{uL,1}$ and $W_{uR,1}$ are unitary matrices that act on the left-
and right-handed up-type quark fields as introduced in
Eq.~(\ref{eq::trafo})
\begin{eqnarray}
  && Q_L\to U_{uL,1} Q_L \,,\quad u_R\to W_{uR,1} u_R\,,
  \label{eq::trafo_QL}
\end{eqnarray} 
and $D_{u,1}$ is a diagonal matrix with positive eigenvalues. Then
\begin{eqnarray}
  Y^u_1=  U_{uL,1} D_{u,1}  W_{uR,1}^\dagger \,,\quad \mbox{with}\quad
  D_{u,1}=\mathrm{diag}(y_{u,1},y_{c,1},y_{t,1})\,,
\end{eqnarray}  
where the diagonal elements of $D_{u,1}$ are the physical couplings
and correspond to the positive square roots of the eigenvalues
of $Y^u_1Y^{u \dagger}_1$.  We can define the unitary matrices $U_{dL}$ and
$W_{dR}$ in a similar way and decompose $Y^d_1$ as
\begin{eqnarray}
  Y^d_1=  U_{dL,1} D_{d,1}  W_{dR,1}^\dagger \,,\quad \mbox{with}\quad
  D_{d,1}=\mathrm{diag}(y_{d,1},y_{s,1},y_{b,1})\,,
\end{eqnarray}  
i.e., $Y^d_1$ is diagonalized via
\begin{eqnarray}
  && Q_L\to U_{dL,1} Q_L \,,\quad d_R\to W_{dR,1} d_R\,.
  \label{eq::trafo_dR}
\end{eqnarray}
However, Eqs.~\eqref{eq::trafo_QL} and \eqref{eq::trafo_dR} are in conflict with each other
and only one of the Yukawa matrices can be diagonalized.
This leads to the definition of the CKM matrix, which, in the basis where the up-type Yukawa matrix
is diagonal, is given by $V=U_{uL,1}^\dagger U_{dL,1}$.
Note that $(2n -1)\stackrel{n=3}{=}5$ unphysical phases can be eliminated from $V$
via further quark field redefinitions.

The discussion up to now is in analogy to the SM.
Within a general 2HDM the unitary transformations discussed above do
not necessarily simultaneously diagonalize the other two Yukawa
matrices $Y^u_2$ and $Y^d_2$. We can still define the additional set
of (non physical) Yukawa couplings
as the positive square roots of the eigenvalues of the matrices
\begin{align}
  M_{u,2}&=Y^u_2Y^{u \dagger}_2= U_{uL,2} D_{u,2}^2  U_{uL,2}^\dagger\quad
  &\mbox{with} \quad D_{u,2}=\mathrm{diag}(y_{u,2},y_{c,2},y_{t,2})\,,\nonumber\\
  M_{d,2}&=Y^d_2Y^{d \dagger}_2= U_{dL,2} D_{d,2}^2  U_{dL,2}^\dagger\quad
  &\mbox{with} \quad D_{d,2}=\mathrm{diag}(y_{d,2},y_{s,2},y_{b,2})\,.
\end{align} 
To summarize, using the unitary rotations in Eq.~(\ref{eq::trafo_QL})
the set of Yukawa matrices transform as
\begin{eqnarray}
  && Y^u_1\to D_{u,1}\,,\qquad\qquad\qquad\quad~~ Y^d_1\to V^\dagger
  D_{d,1}\,,
  \nonumber\\
  && Y^u_2\to N_u=U_{uL,1}^\dagger Y^u_2 W_{uR,1}\,,\quad Y^d_2\to N_d=
  V^\dagger U_{dL,1}^\dagger Y^u_2 W_{dR,1}\,,
  \label{eq::Y1Y2}
\end{eqnarray} 
where $N_u$ and $N_d$ are complex $3\times 3$
matrices. Note that for the special case of a 2HDM with a
$\mathbb{Z}_2$-symmetry only two of the four matrices in
Eq.~(\ref{eq::Y1Y2}) are non-zero. Their eigenvalues define the
physical parameters and their mixing 
matrix is defined in analogy to the CKM matrix in the SM.

We want to stress that within the SM and the four
$\mathbb{Z}_2$-symmetric 2HDMs (cf. Tab.~\ref{tab::types}) the physical Yukawa couplings are
defined as eigenvalues of the Yukawa matrices and thus, by
construction, are invariant under quark flavour
transformations. However, in a general 2HDM only appropriate linear
combinations of the eigenvalues of the Yukawa matrices become
invariant under unitary transformations of the scalar fields and can be interpreted
as physical Yukawa couplings.

The strategy to construct flavour invariants in the Yukawa sector
consists in taking products of Yukawa matrices, contracting over the
internal flavour indices, and taking the trace over the external
flavour indices.  For example, the simplest flavour invariants that
can be constructed within a 2HDM read
\begin{eqnarray}
  I_u^{(1)}=\mathrm{Tr}(Y^u_1Y^{u \dagger}_1+Y^u_2Y^{u \dagger}_2)\,,\quad
  I_d^{(1)}=\mathrm{Tr}(Y^d_1Y^{d \dagger}_1+Y^d_2Y^{d \dagger}_2)\,,
\end{eqnarray}  
where $\mathrm{Tr}$ denotes the trace over the open indices of
the left-handed fermions  $Q_L$. In a generic 2HDM the matrices
\begin{eqnarray}
  M_u = Y^u_1Y^{u \dagger}_1+Y^u_2Y^{u \dagger}_2\quad \mbox{and}\quad 
  M_d= Y^d_1Y^{d \dagger}_1+Y^d_2Y^{d \dagger}_2
  \label{eq::MuMd}
\end{eqnarray}  
are invariant under scalar flavour transformations and one can
thus construct other nine flavour invariants similar to those
for the SM~\cite{Jenkins:2009dy,Feldmann:2015nia}.
Using Section 3.1 of~\cite{Feldmann:2015nia} and adapting the notation (i.e. replacing $U$ and $D$ by $M_u$ and $M_d$) leads to
\begin{alignat}{3}
  &I_{1} ~= \mathrm{Tr}(M_u),~&&I_{3} ~= \mathrm{Tr}(\tilde{M}_u),~&&I_{6} = \mathrm{det}(M_u)~,\nonumber\\
  &I_{2} ~= \mathrm{Tr}(M_d),~&&I_{4} ~= \mathrm{Tr}(\tilde{M}_d),~&&I_{8} = \mathrm{det}(M_d)~,\nonumber\\
  &I_{5} ~= \mathrm{Tr}(M_u M_d),~&&I_{7} ~= \mathrm{Tr}(M_d\,\tilde{M}_u),~&&I_{9} = \mathrm{Tr}(M_u\,\tilde{M}_d)~,\nonumber\\
  &I_{10} = \mathrm{Tr}(\tilde{M}_u\,\tilde{M}_d),~&&I_{11} = -\frac{3i}{8}\mathrm{det}(\left[M_u, M_d\right])~,&&
  \label{eq::SMinv}
\end{alignat}
where $\tilde{M} = M^{-1}\mathrm{det}(M)$.
All $\mathbb{Z}_2$-symmetric 2HDMs have the same eleven invariants as the SM.

In a generic 2HDM further higher rank invariants can be constructed using
tensorial properties of the Yukawa matrices. For example, the
simplest additional type of rank four invariant tensors are
\begin{eqnarray}
  T^{(2)}_{uu}& =& \sum_{a,b}^{1,2}\mathrm{Tr}(Y^u_aY^{u
   \dagger}_b)\mathrm{Tr}(Y^u_bY^{u \dagger}_a)\,,
\label{eq:T2uu}
\end{eqnarray} 
and
\begin{eqnarray}
 T^{(2)}_{dd}\,\, =\,\, \sum_{a,b}^{1,2}\mathrm{Tr}(Y^d_aY^{d
   \dagger}_b)\mathrm{Tr}(Y^d_bY^{d \dagger}_a)\,,
\label{eq:T2dd}
\end{eqnarray}
and similar ones where $\mathrm{Tr}$ is replaced by the determinant.  A
systematic analysis of all independent invariants for a general 2HDM
is, however, beyond the scope of this article.

Let us also mention at this point that the definitions for the physical Yukawa
couplings and mixing matrices introduced above holds to all orders
in perturbation theory.


\section{\label{sec:tech}Technicalities}

In this work we compute the beta functions of the three gauge couplings and
the Yukawa matrices in the $\overline{\rm MS}$
scheme. 

Our calculation of the beta functions are based on the Lagrange
densities in Eqs.~\eqref{eq::yuk} and \eqref{eq::V}.  The
specification to the types I, II, X, and Y is straightforward.
Note that the $\overline{\rm MS}$ renormalization constants
can be computed in the unbroken
phase since they do not depend on the particle masses.

It is convenient to denote the gauge couplings by
$\alpha_1$, $\alpha_2$ and $\alpha_3=\alpha_s$, 
where $\alpha_i = g_i^2/(4\pi)$ and $Y^{f}_{a}$
with $f=u,d,l$ and $a=1,2$ (labeling the scalar doublets).
Furthermore, we introduce $\hat\lambda_{ab,cd}=\lambda_{ab,cd}/(4\pi)$ ($a,b,c,d=1,2$), where
$\lambda_{ab,cd}$ are the quartic coupling in the scalar potential. We define the beta functions via
\begin{eqnarray}
  \mu^2\frac{d}{d\mu^2}\frac{\alpha_i}{\pi} &=& \beta_i(\{\alpha_j, Y^{f}_{a}, \lambda_{ab,cd}\},\epsilon)
  \,, \nonumber\\
  \mu\frac{d}{d\mu} Y^{f}_{a} &=& \beta_{f,a}(\{\alpha_j, Y^{f}_{a}, \lambda_{ab,cd}\},\epsilon)
  \,, \nonumber\\
  \mu^2\frac{d}{d\mu^2} \frac{\hat\lambda_{ab,cd}}{\pi} &=& \beta_{ab,cd}(\{\alpha_j, Y^{f}_{a}, \lambda_{ab,cd}\},\epsilon)
  \,,
\label{eq::beta_fc}
\end{eqnarray}
where $\epsilon=(4-d)/2$. Note that the dependence of the couplings
on the renormalization scale is suppressed. The equations defining the beta function for Yukawa matrices have
to be understood as matrix equations in flavour space.
The gauge couplings are related to the fine structure constant, the weak
mixing angle and the strong coupling as follows
\begin{align}
  \begin{split}
    \alpha_1 &= \frac{5}{3}\frac{\alpha_{\mathrm{QED}}}{\cos^2\theta_W}\,,\\
    \alpha_2 &= \frac{\alpha_{\mathrm{QED}}}{\sin^2\theta_W}\,,\\
    \alpha_3 &= \alpha_s\,,
  \end{split}
  \label{eq::alpha_123}
\end{align}
where the SU(5) normalization has been adopted 
which leads to the factor $5/3$ in the definition of $\alpha_1$.
For models where the first and second generation Yukawa couplings are
neglected, it is convenient to introduce $\alpha_{f,i}=y^2_{f,i}/(4 \pi)$ with
$f=b,t,\tau$ and $i=1,2$.

The beta functions are obtained from the renormalization constants
relating bare and renormalized couplings. For the gauge couplings we have
\begin{eqnarray}
  \alpha_i^{\text{bare}} &=&
  \mu^{2\epsilon}Z_{\alpha_i}(\{\alpha_j\},\epsilon)\alpha_i
  \,.
  \label{eq::alpha_bare}
\end{eqnarray}
From this equation one obtains the following explicit formula
for the beta functions after taking into account that the $\alpha_i^{\text{bare}}$ 
do not depend on $\mu$
\begin{eqnarray}
  \label{eq::renconst_beta}
  \beta_i &=& 
  -\left[\epsilon\frac{\alpha_i}{\pi}
    +\frac{\alpha_i}{Z_{\alpha_i}}
    \sum_{{j=1},{j \neq i}}
    \frac{\partial Z_{\alpha_i}}{\partial \alpha_j}\beta_j\right]
  \left(1+\frac{\alpha_i}{Z_{\alpha_i}}
    \frac{\partial Z_{\alpha_i}}{\partial \alpha_i}\right)^{-1}
  \,.
\end{eqnarray}
The first term in the first factor of Eq.~\eqref{eq::renconst_beta} originates
from the term $\mu^{2\epsilon}$ in Eq.~\eqref{eq::alpha_bare} and vanishes in
four space-time dimensions. Equations ~\eqref{eq::alpha_bare} and ~\eqref{eq::renconst_beta}
hold for the Yukawa couplings  only for models where the
Yukawa matrices are diagonal, e.g., in case only the third
generation Yukawa couplings are taken into account.

The generalization of Eq.~\eqref{eq::renconst_beta} to incorporate
  tensor-like couplings, like the Yukawa matrices and quartic couplings, is
  straightforward to derive. However, care has to be taken when computing
  derivatives of renormalization constants.  Furthermore, the relations
  between Yukawa matrix and quartic coupling beta functions and the
  corresponding renormalization constants take a slightly different form than
  in Eq.~\eqref{eq::renconst_beta}, since in general, due to the tensorial
  nature, it is not possible to compute the inverse of the renormalization
  constants.  For more details see Ref.~\cite{Herren:2017}.  

Another option would be to derive the scale dependence of the eigenvalues
  of the Yukawa matrices and quartic couplings starting from the definition in
  Eq.~\eqref{eq::beta_fc}. 

Note that the one-loop results of $Z_{\alpha_i}$ only contain $\alpha_i$,
whereas at two loops all other couplings are present except for the quartic
couplings.  The renormalization constants of the Yukawa matrices contain all
couplings except the quartic couplings already at one-loop order, while the
quartic couplings enter at two loops.  Therefore, it is necessary to compute
the renormalization constants and beta functions of the quartic couplings only
at one-loop order.\footnote{Our results can be found in the ancillary
    files to this paper~\cite{progdata}.}.

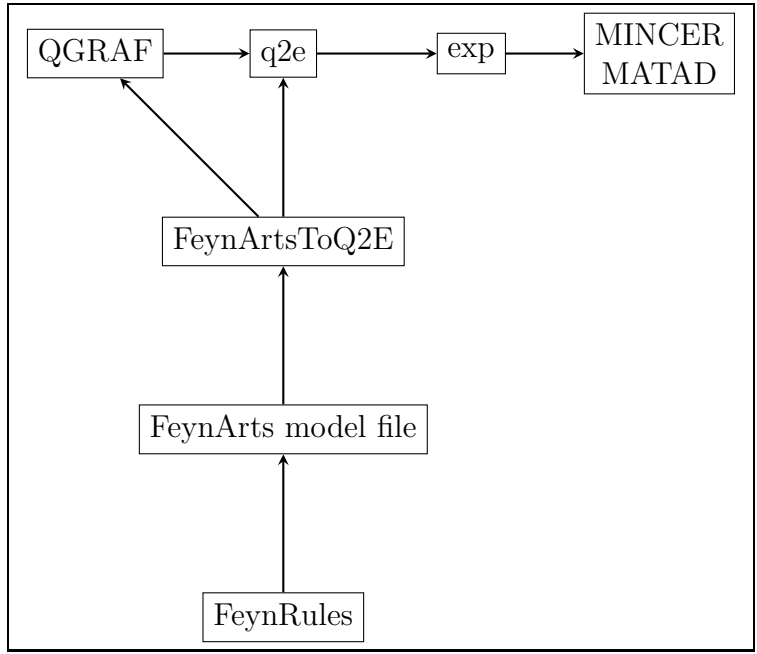
\begin{figure}[t]
  \begin{center}
    \fcolorbox{black}{white}{
\begin{tikzpicture}[node distance=2.5cm, every text node part/.style={align=center}]
\node (QGRAF) [rec] {QGRAF};
\node (q2e) [rec, right of=QGRAF] {q2e};
\node (exp) [rec, right of=q2e] {exp};
\node (MINCER) [rec, right of=exp] {MINCER\\MATAD};
\node (F2Q) [rec, below of=q2e] {FeynArtsToQ2E};
\node (FeynArts) [rec, below of=F2Q] {FeynArts model file};
\node (FeynRules) [rec, below of=FeynArts] {FeynRules};
\draw [arrow] (QGRAF) -- (q2e);
\draw [arrow] (q2e) -- (exp);
\draw [arrow] (exp) -- (MINCER);
\draw [arrow] (F2Q) -- (q2e);
\draw [arrow] (F2Q) -- (QGRAF);
\draw [arrow] (FeynArts) -- (F2Q);
\draw [arrow] (FeynRules) -- (FeynArts);
\end{tikzpicture}
    }
    \caption{\label{fig::setup}Flowchart illustrating the workflow used for the
      calculation of the two- and three-point functions.}
  \end{center}
\end{figure}

For our calculation we use the automated setup developed for the calculation
of the SM gauge beta functions to three
loops~\cite{Mihaila:2012fm,Mihaila:2012pz}. For convenience we repeat the
flowchart which illustrates the interaction of the various program packages
in Fig.~\ref{fig::setup}.

In a first step we implement the unbroken version of the general 2HDM
discussed in Section~\ref{sec:2HDM} in the package
\verb|FeynRules|~\cite{Christensen:2008py} which generates a model file
for~\verb|FeynArts|~\cite{Hahn:2000kx}. The program
\verb|FeynArtsToQ2E|~\cite{diss_salomon} works on the model file and
translates it into input files for \verb|QGRAF|~\cite{Nogueira:1991ex} and
\verb|q2e|~\cite{Harlander:1997zb,Seidensticker:1999bb,q2eexp}.  \verb|QGRAF|
is used for the generation of the amplitudes which are translated by
\verb|q2e| and \verb|exp|~\cite{Harlander:1997zb,Seidensticker:1999bb,q2eexp}
to \verb|FORM|~\cite{Kuipers:2012rf} code. The latter is processed by
\verb|MINCER|~\cite{Larin:1991fz} and/or
\verb|MATAD|~\cite{Steinhauser:2000ry} which compute the Feynman integrals and
outputs the $\epsilon$ expansion of the result.

For the first part of the calculation up to the generation of the input files for
\verb|QGRAF| and \verb|q2e| no parallelization is necessary. The individual
steps take at most a few minutes. However, the parallelization of the
horizontal part of the flowchart (cf. Fig.~\ref{fig::setup}) is essential since
for some of the Green's function we have to deal with several hundred
thousands of diagrams. Once \verb|QGRAF| has produced the output file all
following steps can be applied in parallel to blocks of diagrams which
typically contain 1000 Feynman amplitudes.

\begin{table}[t] {\begin{center} \begin{tabular}[t]{c||r|r|r} \hline
        \multicolumn{4}{c}{2-point functions} \\
        \hline
        \# loops & 1 & 2 & 3 \\
        \hline
        $BB$                    &   16 & 450 &  49\,256 \\
        $W_3W_3$                &   19 & 534 &  57\,665 \\
        $gg$                    &    9 & 170 &  13\,671 \\
        $c_g \bar{c}_g$         &    1 &  12 &      447 \\
        $c_{W_3} \bar{c}_{W_3}$ &    2 &  46 &   2\,880 \\
        $q \bar{q}$             &   11 & 659 &  75\,980 \\
        $l \bar{l}$             &   10 & 567 &  63\,853 \\
        $\Phi^{0*}_1 \Phi^0_1$  &   8 & 436 &  47\,613 \\
        $\Phi^{0*}_1 \Phi^0_2$  &   4 & 224 &  28\,648 \\
        $\Phi^{0*}_2 \Phi^0_1$  &   4 & 224 &  28\,648 \\
        $\Phi^{0*}_2 \Phi^0_2$  &   8 & 436 &  47\,613 \\
        \hline \end{tabular} \hspace*{2em} \begin{tabular}[t]{c||r|r|r} \hline
        \multicolumn{4}{c}{3-point functions} \\
        \hline
        \# loops & 1 & 2 & 3 \\
        \hline
        $BBB$                       &   44 & 2\,472 & 401\,460 \\
        $c_g \bar{c}_g g$           &    2 &     66 &   3\,722 \\
        $c_{W_1} \bar{c}_{W_2} W_3$ &    2 &    117 &  11\,849 \\
        $\Phi^0_1\Phi^{0*}_1 W_3$   &   22 & 2\,538 & 417\,759 \\
        $\Phi^0_1\Phi^{0*}_2 W_3$   &   12 & 1\,274 & --- \\
        $\Phi^0_2\Phi^{0*}_1 W_3$   &   12 & 1\,274 & --- \\
        $\Phi^0_2\Phi^{0*}_2 W_3$   &   22 & 2\,538 & --- \\
        $d\bar{d}\Phi^0_i$          &   17 & 2\,622 & 493\,742 \\
        $u\bar{u}\Phi^{0*}_i$       &   17 & 2\,622 & 493\,742 \\
        $l\bar{l}\Phi^0_i$          &   16 & 2\,337 & 426\,741 \\
        \hline \end{tabular} \caption[]{\label{tab::GFs}The number of Feynman
        diagrams contributing to the one-, two- and three-loop Green's
        functions evaluated in this work.  We computed the Yukawa vertices for
        both $\Phi_1$ and $\Phi_2$.  Note that for some Green's functions less
        diagrams had to be calculated than in \cite{Mihaila:2012pz} since we
        only considered one fermion generation with matrix-like Yukawa
        couplings. We did not compute three-loop corrections 
        to the vertices $\Phi^0_i\Phi^{0*}_j W_3$ with $ij=12,21$ and $22$.
        The corresponding two-loop results are needed for the
        three-loop calculation of $\Phi^0_1\Phi^{0*}_1 W_3$.}
  \end{center}
}
\end{table}

We perform the calculation in Lorenz gauge using general
gauge parameters for each gauge group. It is an important cross check that
they drop out in the expressions for the renormalization constants (and beta
functions) of the gauge and Yukawa couplings.

\begin{figure}[t]
  \begin{center}
    \includegraphics[width=1\textwidth]{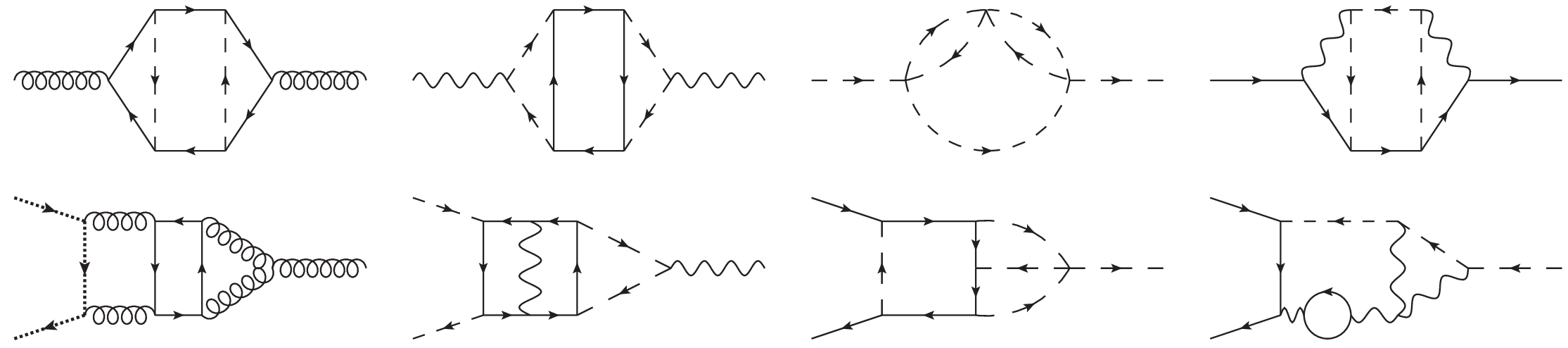}
    \caption{\label{fig::diags}Sample Feynman diagrams contributing to 
      the Green's functions which have been used for our calculation of the 
      gauge and Yukawa coupling renormalization constants.
      Solid, dashed, dotted, curly and wavy lines denote fermions,
      scalar bosons, ghosts, gluons and electroweak gauge bosons,
      respectively.} 
  \end{center}
\end{figure}

The described setup is used to compute various $\overline{\rm MS}$
renormalization constants for fields and vertices. They are required for the
construction of the renormalization constants for the gauge, Yukawa and quartic
couplings. 

For the SM and
2HDM with $\mathbb{Z}_2$-symmetry, one can perform the calculation in a basis
where all the Yukawa matrices are diagonal and the elements of the CKM matrix are present only
in the vertices containing charged currents. In such a basis, the Lagrangian
parameters are  physical parameters and the number of
free parameters is reduced to the number of independent degrees of freedom.\\   
In Table~\ref{tab::GFs} we list all Green's functions, which we have considered
in the course of the calculations performed in this paper, and the number of
generated Feynman amplitudes up to three loops. We used the following notation for the fields:
$B$ and $W_i$ denote the gauge bosons, $c_x$
refers to the ghost fields and $\Phi_i^0$ and $\Phi_i^\pm$ ($i=1,2$) are the
neutral and charged components of the scalar doublets. In Fig.~\ref{fig::diags} we show
typical Feynman diagrams contributing to the individual Green's functions.

Due to Ward identities there are various choices for each
gauge coupling to obtain
$Z_{\alpha_i}$:
\begin{eqnarray}
  Z_{\alpha_1} &=& \frac{1}{Z_{BB}}
  \,\,=\,\,
  \frac{(Z_{f_L\bar{f}_LB})^2} {Z_{BB} Z_L^{f} Z_L^{f}}
  \,\,=\,\, \ldots
  \,,\nonumber\\
  Z_{\alpha_2} 
  &=& 
  \frac{(Z_{c_{W_1} \bar{c}_{W_2} W_3})^2} {Z_{W_3W_3} (Z_{c_{W_3} \bar{c}_{W_3}})^2} 
  \,\,=\,\,
  \frac{(Z_{W_1W_2W_3})^2} {(Z_{W_3W_3})^3} 
  \,\,=\,\,
  \frac{(Z_{f_L\bar{f}_L W_3})^2} {Z_{W_3W_3} Z_L^{f} Z_L^{f}}
  \nonumber\\&=&
  \frac{(Z_{\Phi^+_1\Phi^-_1 W_3})^2} {Z_{W_3W_3} (Z_{\Phi^+_1\Phi^-_1})^2}
  \,\,=\,\,
  \frac{(Z_{\Phi^0_1\Phi^0_1 W_3})^2} {Z_{W_3W_3} (Z_{\Phi^0_1\Phi^0_1})^2}
  \,\,=\,\, \ldots
  \,,\nonumber\\
  Z_{\alpha_3} 
  &=& 
  \frac{(Z_{c_{g} \bar{c}_{g} g})^2} {Z_{gg} (Z_{c_{g} \bar{c}_{g}})^2} 
  \,\,=\,\,
  \frac{(Z_{ggg})^2} {(Z_{gg})^3} 
  \,\,=\,\,
  \frac{(Z_{f_L\bar{f}_Lg})^2} {Z_{gg} Z_L^{f} Z_L^{f}}
  \,\,=\,\, \ldots
  \,,
\end{eqnarray}
where we have used $Z_{c_{W_1} \bar{c}_{W_1}} = Z_{c_{W_2} \bar{c}_{W_2}} =
Z_{c_{W_3} \bar{c}_{W_3}}$ and $Z_{W_1W_1} = Z_{W_2W_2} = Z_{W_3W_3}$. Here
$Z^f_L$, $Z^f_R$ with\footnote{In the case of $Z_{\alpha_3}$ we have $f=u,d$.}
$f=u,d,l$ stand for the wave functions renormalization of the left- and right
handed fermion fields. Their explicit definition will be introduced in the
next section.

The lowest number of Feynman diagrams are generated for the Green's functions
involving ghosts. Thus, our default choice for the computation of the gauge
coupling renormalization constants are the gauge boson-ghost vertices and the
corresponding two-point functions. Other vertices have been considered
to have powerful cross checks.
Due to the Ward identity the renormalization constant for $\alpha_1$ is
given by the inverse renormalization constant of the $U(1)$ gauge boson propagator.
We have performed an explicit calculation of $Z_{BB}$, $Z_{c_{W_1}
  \bar{c}_{W_2} W_3}$, $Z_{W_3W_3}$, $Z_{c_{W_3} \bar{c}_{W_3}}$,
$Z_{u_L\bar{u}_L W_3}$, $Z_L^{u} Z_L^{u}$, $Z_{\Phi^0_1\Phi^0_1 W_3}$,
$Z_{\Phi^0_1\Phi^0_1}$, $Z_{c_{g} \bar{c}_{g} g} $, $Z_{gg}$, $Z_{c_{g}
  \bar{c}_{g}}$. 

For the Yukawa matrices we are restricted to vertices which involve
components of the scalar doublets as well as left- and right-handed fermion
fields.  The explicit definition of the Yukawa matrix renormalization
constants will be postponed to the next section.

At the end of this section a comment concerning $\gamma_5$ is in order.
For the computation of some of the Green's functions an odd number of
$\gamma_5$ matrices is present in the traces. We have checked that it is
sufficient to follow the prescription provided in Ref.~\cite{Mihaila:2012pz}
in the context of the SM. This means that a formal replacement of expressions like
\begin{align}
  \mathrm{Tr}\left(\gamma^\mu\gamma^\nu\gamma^\rho\gamma^\sigma\gamma_5\right)
  = -4\mathrm{i}\tilde{\epsilon}^{\mu\nu\rho\sigma} + \mathcal{O}(\epsilon)~.
  \label{eq:epstilde}
\end{align}
is applied, where $\tilde{\epsilon}^{\mu\nu\rho\sigma}$ is antisymmetric
in all indices. In practice the product of two such objects occurs,
where all indices are contracted, which we replace by
\begin{align}
  \tilde{\epsilon}^{\mu\nu\rho\sigma}\tilde{\epsilon}_{\mu'\nu'\rho'\sigma'} =
  g^{[\mu\phantom{]}}_{[\mu'\phantom{]}}g^{\phantom{[}\nu\phantom{]}}_{\phantom{[}\nu'\phantom{]}}
  g^{\phantom{[}\rho\phantom{]}}_{\phantom{[}\rho'\phantom{]}}
  g^{\phantom{[}\sigma]}_{\phantom{[}\sigma']}~.
\end{align} 
The square brackets denote complete antisymmetrization. This leads to
the correct result in the limit $d \rightarrow 4$. We have checked explicitly
that the ambiguity of $\mathcal{O}(\epsilon)$ in Eq.~(\ref{eq:epstilde}) is
multiplied by at most simple poles in $\epsilon$ and therefore does not lead to
ambiguous renormalization constants and beta functions.


\section{\label{sec:ren}$\overline{\rm MS}$ renormalization of the general 2HDM}


\subsection{Renormalization constants}

For the computation of the renormalization constants for fields, couplings and
vertices we follow the procedure described in Ref.~\cite{Steinhauser:1998cm}.
However, since we consider general Yukawa couplings which are non-diagonal
both in flavour as well as in doublet space, several modifications have to be
applied, in particular for the calculation of the fermion and scalar wave
function renormalization constants, and the renormalization constants for the
Yukawa matrices. These issues are discussed in this section.

The renormalized inverse fermion propagator can be written as
\begin{align}
  S_F^{-1}(p) = \slashed{p}\left[\mathrm{P}_L\left(\sqrt{Z_L^f}\right)^\dagger
    (1 +\Sigma_L(p^2))\sqrt{Z_L^f}
    +\mathrm{P}_R\left(\sqrt{Z_R^f}\right)^\dagger(1 +
    \Sigma_R(p^2))\sqrt{Z_R^f}\right]\,,
  \label{eq:fermionprop}
\end{align}  
where $P_{L/R} = (1\mp\gamma_5)/2$, $\Sigma_{L/R}$ are the left- and
right-handed parts of the fermion self energy and $Z_{L/R}^f$ are the
renormalization constants for the left- and right-handed fermion
fields. Both $\Sigma_{L/R}$ and $Z_{L/R}^f$ are matrices in flavour space
where flavour indices have been suppressed.
The index $f\in\{u,d,l\}$ indicates whether the up-, down- or lepton matrix
shall be considered.

The renormalized two-point function of the scalar fields can be written as
\begin{align}
  \Pi^{\rm ren}(p^2) = \left(\sqrt{Z^\Phi}\right)^\dagger\Pi(p^2)\sqrt{Z^\Phi}
  \,,
  \label{eq:scalarprop}
\end{align}  
where the wave function renormalization constant $Z^\Phi$ and $\Pi(p^2)$
are matrices in doublet space. The corresponding indices have been
suppressed.  In $\mathbb{Z}_2$-symmetric models $Z^\Phi$ has to be
diagonal, which we have checked up to three-loop level.

From Eqs.~\eqref{eq:fermionprop} and \eqref{eq:scalarprop} the following relations can be derived:
\begin{eqnarray}
  Z_L^f & = & 1 - K_\epsilon\left[\left(\sqrt{Z_L^f}\right)^\dagger
    (1 + \Sigma_L(p^2))\sqrt{Z_L^f}\right]
  \,,  \nonumber\\
  Z_R^f & = & 1 - K_\epsilon\left[\left(\sqrt{Z_R^f}\right)^\dagger
    (1 + \Sigma_R(p^2))\sqrt{Z_R^f}\right]
  \,,  \nonumber\\
  Z^\Phi & = & 1 -
  K_\epsilon\left[\left(\sqrt{Z^\Phi}\right)^\dagger\Sigma(p^2)\sqrt{Z^\Phi}\right]
  \label{eq::propren}\,.
\end{eqnarray}
The operator $K_\epsilon$ extracts the poles in $\epsilon$. Solving these
equations recursively allows to determine the corresponding renormalization
constants. Let us stress at this point that from the equations above we can
compute only the hermitian parts of the renormalization matrices $Z_L^f,
Z_R^f$ and $Z^\Phi$. In the SM the anti-hermitian parts of
the quark wave functions renormalizations are related to the renormalization
of the CKM matrix~\cite{Gambino:1998ec,Balzereit:1998id,Denner:2004bm}.
In the next section we will also introduce anti-hermitian contributions to the
renormalization matrices defined above. However, in our case, they
should not be identified with the renormalization of any physical quantity.

Let us in a next step use this information to obtain formulae which allow to
compute the renormalization constants for the scalar-fermion vertices and the
Yukawa couplings. The Yukawa vertex renormalization constants for a fermion
of type $f$ can be extracted from
\begin{align}
   \sum_{\beta}\sum_{b=1}^2 Z^{ff\Phi}_{ab,\alpha\beta}
Y^f_{b,\beta\alpha^\prime}
   = Y^f_{a,\alpha\alpha^\prime} - K_\epsilon\left[
     \sum_{\beta,\gamma} \sum_{b=1}^2
     \left( \sqrt{Z_L^f}_{\alpha\beta}\right)^\dagger
     \sqrt{Z^\Phi}_{ab} \Gamma(p,0)_{b,\beta\gamma}
     \sqrt{Z_R^f}_{\gamma\alpha^\prime}
   \right]\,,
   \label{eq::ZfYuk}
\end{align}
where for convenience the flavour ($\alpha,\beta,\gamma$) and the doublet
($a,b$) indices are shown explicitly. The sums over $\beta$ and $\gamma$ run
over all down-(up-)type fermions in case $\alpha$ is a down-(up-)type fermion.
Note that in Eq.~(\ref{eq::ZfYuk}) $\Gamma(p,0)_{b,\beta\gamma}$ is the
vertex function where one of the external momenta is set to zero and
the external fields are $\Phi_b$ and fermions with flavour $\beta$ and
$\gamma$. Furthermore, the Yukawa coupling in the tree-level contribution
of $\Gamma(p,0)_{b,\beta\gamma}$ is not renormalized.

Once $\sum_{\beta}\sum_{b=1}^2 Z^{ff\Phi}_{ab,\alpha\beta}
Y^f_{b,\beta\alpha^\prime} $ is obtained from Eq.~(\ref{eq::ZfYuk})
the Yukawa matrix renormalization constants ($\Delta Y^f_a$) can be
computed from
\begin{align}
   \sum_{\beta}\sum_{b=1}^2 Z^{ff\Phi}_{ab,\alpha\beta} Y^f_{b,\beta\alpha^\prime}
   = \sum_{\beta,\gamma}\sum_{b=1}^2
   \left(\sqrt{Z_L^f}_{\alpha\beta}\right)^\dagger
   \sqrt{Z^\Phi}_{ab}
   (Y^f_b+\Delta Y^f_b)_{\beta\gamma}
   \sqrt{Z_R^f}_{\gamma\alpha^\prime}
   \,.
   \label{eq::ZfYuk_2}
\end{align}
This equation has to be solved iteratively for $(Y^f_a+\Delta Y^f_a)$.


\subsection{Invariants in the quark sector}

The renormalization constants introduced in Eq.~(\ref{eq::propren}) are used
to derive the corresponding anomalous dimensions.  Note, however, that the
anomalous dimensions might contain poles in $\epsilon$.  This is not surprising, since in
the case of general Yukawa matrices we do not take into account the invariance
of the theory under unitary rotations like those given in
Eq.~(\ref{eq::trafo}). In other words, from the $72$ parameters\footnote{4
  complex $3\times 3$ matrices with 18 parameters each.} of the Yukawa sector
of the general 2HDM $30$ can be eliminated using flavour
transformations~\cite{Santamaria:1993ah}\footnote{The flavour symmetry group
$[U(3)]^3\otimes U(2)$ is broken by the Yukawa sector to $U(1)$, leading to 30 broken generators.}.
In contrast to the case of
Lagrangian parameters, the beta functions of the flavour invariants
(cf. Eq.~\eqref{eq::SMinv}) are finite, because the flavour symmetry
relations are by construction taken into account in such quantities.

We also want to remark that the gauge couplings are trivially invariant under
unitary flavour transformations and
the corresponding beta functions do not suffer from uncanceled singularities.
On the other hand, in analogy to the Yukawa matrices,
we expect that in the case of the quartic couplings of
the scalar potential only certain combinations of them have finite beta functions.

In the SM and in $\mathbb{Z}_2$-symmetric 2HDMs there are eleven flavour
invariants in the quark sector as has been discussed in Section~\ref{sec:2HDM}. From them the six quark masses, three CKM
mixing angles, and the cosine and the sign of the CP-violating phase can be
derived.  Their behaviour under renormalization group evolution has been
studied up to two loops, for example, in Ref.~\cite{Feldmann:2015nia}.  We
have extended the analysis and checked by an explicit calculation that all
eleven invariants in Eq.~(\ref{eq::SMinv}) of the SM and
$\mathbb{Z}_2$-symmetric 2HDMs have finite anomalous dimensions at three
loops.  From the three-loop
anomalous dimensions of the eleven invariants mentioned above one can derive
the beta functions for the physical
couplings and the CKM mixing angles to the same order. 

As already mentioned in Section~\ref{sec:2HDM}, we have not classified
all flavour invariants in the general 2HDM.  However, an explicit
calculation for the invariants
\begin{align}
&\mathrm{Tr}\left(Y^u_a Y^{u\dagger}_a\right),
\mathrm{Tr}\left(Y^d_a Y^{d\dagger}_a\right),
\mathrm{Tr}\left(Y^u_a Y^{u\dagger}_b Y^u_b Y^{u\dagger}_a\right),
\mathrm{Tr}\left(Y^u_a Y^{u\dagger}_a Y^u_b Y^{u\dagger}_b\right),
\nonumber\\&
\mathrm{Tr}\left(Y^d_a Y^{d\dagger}_b Y^d_b Y^{d\dagger}_a\right),
\mathrm{Tr}\left(Y^d_a Y^{d\dagger}_a Y^d_b Y^{d\dagger}_b\right),
\mathrm{Tr}\left(Y^u_a Y^{u\dagger}_a Y^d_b Y^{d\dagger}_b\right),
\mathrm{Tr}\left(Y^u_a Y^{u\dagger}_b Y^d_a Y^{d\dagger}_b\right),
\nonumber\\&
\mathrm{Tr}\left(Y^u_a Y^{u\dagger}_b\right) \mathrm{Tr}\left(Y^u_b
Y^{u\dagger}_a\right), \mathrm{Tr}\left(Y^d_aY^{d\dagger}_b\right)
\mathrm{Tr}\left(Y^d_b Y^{d\dagger}_a\right),
\mathrm{Tr}\left(Y^u_a Y^{u\dagger}_b\right) \mathrm{Tr}\left(Y^d_a
Y^{d\dagger}_b\right),
\end{align}
where we sum over $a$ and $b$ and the trace is taken over the fermionic indices,
shows that all poles cancel.

\subsection{\label{sec:simplified}Invariants for a simplified model}

In this Subsection we consider a simplified version of the general
2HDM. Explicitly, we study the case where the Yukawa interactions for the
first and second generations are neglected. As a consequence, the Yukawa
matrices reduce to complex numbers, parameterizing the Yukawa couplings for
the $t$ and $b$ quarks and only scalar flavour symmetries
occur. Following Ref.~\cite{Santamaria:1993ah}, one observes that from
the eight parameters\footnote{We have $y_i^f \in \mathbb{C}$ with $f = u,d$ and $i = 1,2$.} in the Yukawa sector
$(n^2-1)\stackrel{n = 2}{=}3$ can be rotated away.  We also notice that the up- and
down-type Yukawa couplings transform as vectors under unitary rotations of the
scalar fields, see Eq.~(\ref{eq::trafo2}) where $U_Q$ and $U_{u,d}$
are replaced by the identity matrix for  this simplified model.
We thus rotate the scalar fields with the following matrix
\begin{eqnarray}
  U_\Phi&=\frac{1}{\sqrt{y_1^t y^{t \ast}_1+y_2^t y^{t \ast}_2}}&\left(\begin{array}{cc}
      y^{t \ast}_1&y^{t \ast}_2\\
      -y^{t}_2&y^{t}_1
    \end{array}
  \right)\,.
\end{eqnarray}
Under this transformation the scalar fields change to 
\begin{eqnarray}
  \Phi_1^\prime &=& \frac{\delta_{i j}  y^{t \ast}_i}{\sqrt{I_t}} \Phi_j\,,\nonumber\\
  \Phi_2^\prime &=& \frac{\varepsilon_{i j} y^{t}_i}{\sqrt{I_t}} \Phi_j
\end{eqnarray} 
with $I_{t}= \delta_{i j} y^{t}_i y^{t \ast}_j$.  $\delta_{a b}$ and
$\varepsilon_{a b}$ denote the Kronecker delta and Levi-Civita tensor,
respectively, and the sum over the repeated indices $i,j=1,2$ is assumed.
Furthermore, the Yukawa couplings transform as
\begin{align}
  &y^{t \prime}_1 = \sqrt{\delta_{i j} y^{t}_i y^{t \ast}_j}\,=\, \sqrt{I_{t}}\,,
  &&y^{t \prime}_2 \,=\, 0\,,
  \nonumber\\
  &y^{b \prime}_1 = \frac{\delta_{i j} y^{t}_i y^{b}_j}{\sqrt{I_t}}\,,
  &&y^{b \prime}_2 \,=\, \frac{\varepsilon_{i j} y^{b}_i y^{t\ast}_j}{\sqrt{I_t}}\,.
  \label{eq::ytprime}
\end{align}
Taking into account the tensorial properties of $\delta_{ab}$ and $\varepsilon_{ab}$ and the
transformations of the Yukawa couplings under unitary rotations of the scalar fields,
one can easily prove that both the rotated fields and couplings are actually
flavour invariants. In other words, in the new basis the Lagrangian parameters
are expressed through flavour invariants and are therefore directly related to
physical quantities.

An explicit calculation shows that the anomalous dimensions of the new fields
$\Phi_a^\prime$ with $a=1,2$ and the beta functions of the new couplings
$y^{t \prime}_1,\,y^{b \prime}_1$ and $y^{b \prime}_2$ are finite
through three loops. This is not the case for the original basis, where the
Yukawa sector contains too many parameters. The new
basis makes use of the flavour symmetries and gets rid of one of the up-type
Yukawa couplings, the other one is rendered real. It is also important
to notice that the relation $y^{t \prime}_2 = 0$ is stable under
renormalization. To verify this statement, we checked through three
loops that the beta functions of the three Yukawa couplings obtained after
the rotation to the new basis can be expressed only in terms of couplings
present in this basis. At this stage, also the scalar quartic couplings have
to be transformed according to Eqs.~(\ref{eq::trafo3}).  This shows that the
set of couplings in the new basis is complete. Even for this simplified
model the explicit three-loop results are quite lengthy. Thus, in
Section~\ref{sec:yuk} we only present results for $\beta_{y^{t\prime}_1}$.

\subsection{Poles in anomalous dimensions}

In this Subsection we describe a practical method, which allows to use the
beta functions and anomalous dimensions for a general 2HDM, although they
develop poles in the first place. A transformation to physical observables,
which, as mentioned above, becomes quite involved, is not necessary.
We follow Ref.~\cite{Bednyakov:2014pia}, where this issue has been discussed for 
the case of the SM. It is argued that the poles 
can be eliminated by choosing the square roots of the renormalization
constants to be non-hermitian. 
We define
\begin{align}
  \sqrt{Z} = \sqrt{Z}_H\sqrt{Z}_U
  \,,
  \label{eq::Z}
\end{align}
where $Z$ is any of the renormalization constants introduced in
Eq.~(\ref{eq::propren}). The subscripts $H$ and $U$ in Eq.~(\ref{eq::Z})
denote the hermitian and unitary parts.

In order to obtain $Y^u + \Delta Y^u$ one has to invert
Eq.~(\ref{eq::ZfYuk_2}). This can be done by either choosing
hermitian square roots or by allowing for additional unitary factors
which leads to the following relation
\begin{align}
  Y'^u_{a} + \Delta Y'^u_{a} = \sqrt{Z_L}_U\left(Y^u_{b} + \Delta
    Y^u_{b}\right)\left(\sqrt{Z_R}\right)^\dagger_U\sqrt{Z_\Phi}_{U,ba} 
  \,,
\end{align}
where the $Y^u + \Delta Y^u$ is calculated with hermitian square roots. This
equation resembles the transformation in Eq.~(\ref{eq::trafo2}) and
therefore a choice of the unitary part of the square root of the $Z$ factors
can be interpreted as a certain choice of basis.

$\sqrt{Z}_H$ is fixed by the poles of the corresponding two point functions
in Eq.~(\ref{eq::propren}) and can be used to obtain the hermitian part of the
corresponding anomalous dimensions, i.e., the combination
$\gamma + \gamma^\dagger$. For the left- and right-handed fermion fields and
the scalar fields considered in Eq.~(\ref{eq::propren}) we observe that
$\gamma + \gamma^\dagger$ is finite whereas the individual terms are not.

Note, however, that $\sqrt{Z}_U$ is an arbitrary unitary matrix
which can be chosen such that $\gamma$ is finite.
This choice is not unique and it is possible to also influence the
finite parts of the anomalous dimensions (and in general the beta functions) this way. We
will postpone the discussion of this apparent ambiguity and its physical
significance to Subsection~\ref{sec::amb} and concentrate in the following
on the discussion of the left-handed quark fields. 

At one-loop order there is no possibility to construct $\sqrt{Z^Q_{L}}_U \neq
\mathbb{I}$ and therefore $\sqrt{Z^Q_{L}}$ is purely hermitian.
At two-loop order there is one unitary combination of Yukawa matrices
\begin{align}
  \sqrt{Z^Q_{L}}_U = \mathbb{I} + 
  \left(\frac{a}{\epsilon^2}+\frac{b}{\epsilon}\right)\left[Y^u Y^{u\dagger},
    Y^d Y^{d\dagger}\right]~,
\end{align}
where $a$ and $b$ are arbitrary constants.  A nonzero value for $b$ enters
into the finite parts of the left-handed quark field anomalous dimension and
therefore into the beta functions for $Y^u$ and $Y^d$, contributing to the
mentioned ambiguity (see below Subsection~\ref{sec::amb}).

One can choose a nonzero value for $a$ to cancel possible $\epsilon$ poles
in the non-hermitian part of the two-loop anomalous dimension. However, such
poles can not appear as can be seen by the following arguments:
For the anomalous dimensions we schematically write
\begin{align}
  \gamma &= -\sqrt{Z^Q_{L}}^{-1}\mu\frac{\mathrm{d}}{\mathrm{d}\mu}\sqrt{Z^Q_{L}}\,,
  \nonumber\\
  \gamma^\dagger &= 
  -\left(\mu\frac{\mathrm{d}}{\mathrm{d}\mu}\sqrt{Z^Q_{L}}^\dagger\right)\sqrt{Z^Q_{L}}^{\dagger
    -1}\,,
\end{align}
where the second equation simplifies to
\begin{align}
  \gamma^\dagger &=
  -\left(\mu\frac{\mathrm{d}}{\mathrm{d}\mu}\sqrt{Z^Q_{L}}\right)\sqrt{Z^Q_{L}}^{-1}
\end{align}
in the case hermitian square roots are chosen. Thus, the
anomalous dimension is hermitian (and finite) if the commutator
\begin{align}
  \left[\left(\mu\frac{\mathrm{d}}{\mathrm{d}\mu}\sqrt{Z^Q_{L}}\right), \sqrt{Z^Q_{L}}^{-1}\right] = 0
  \label{eq::comm}
\end{align}
is satisfied. At two-loop order only the contribution where both terms in
Eq.~(\ref{eq::comm}) receive one-loop contributions involving two Yukawa
matrices could possibly lead to a nonvanishing commutator of the form
\begin{align}
  \left[Y^u Y^{u\dagger}, Y^d Y^{d\dagger}\right]~.
  \label{eq::[YY,YY]}
\end{align}
However, an explicit calculation in the general 2HDM model shows that $Y^u
Y^{u\dagger}$ and $Y^d Y^{d\dagger}$ appear with the same coefficients
in the renormalization constant and thus the commutator is zero and no
non-trivial factor $\sqrt{Z^Q_{L}}_U$ is needed.

At three loops we have $\gamma \neq \gamma^\dagger$ in case hermitian square
root factors are chosen. For example, there are contributions to $Z^Q_{L}$
involving $Y^u Y^{u\dagger}$ or $Y^d Y^{d\dagger}$, however,
with different prefactors for up- and down-type quarks due to the presence of
hyper-charge contributions. This leads to a non-vanishing commutator in
Eq.~(\ref{eq::comm}). Thus, the necessity to choose a nontrivial factor 
$\sqrt{Z^Q_{L}}_U$ arises from three loops.

Similar arguments hold for the scalar and right-handed quark fields. 


\subsection{\label{sec::amb}Ambiguities in the Yukawa matrix beta function}

The possibility to introduce $\sqrt{Z}_U\not=\mathbb{I}$ introduces an
ambiguity in the definition of the renormalization constants, anomalous
dimensions and beta functions.  Nevertheless, let us stress that this
statement only holds for the unphysical parameters, e.g. the Yukawa
matrices. Once we construct flavour invariants the unitary roots cancel and
their anomalous dimensions are finite and unambiguous. Consequently, the
anomalous dimensions of the physical quantities derived from them are finite
and unambiguous, as expected.

We verified the cancellation of the unitary roots and consequently the poles
in the beta functions for all eleven invariants
of the quark sector of $\mathbb{Z}_2$-symmetric 2HDMs. Furthermore, we checked
the cancellation in the general 2HDM for the invariants introduced in
Section~\ref{sec:2HDM} as well as further invariants entering the gauge
coupling beta functions, which we will present in the next section (see Eqs.~\eqref{eq::m2} and \eqref{eq::t2}).

In addition, we also performed numerical checks by computing the Yukawa
matrices in $\mathbb{Z}_2$-symmetric 2HDMs at a low scale and run them up to
$10^{16}$~GeV for different choices of $\sqrt{Z}_U$, modifying the finite part
of the beta functions.  While the Yukawa matrices themselves differ at the
high scale, depending on the choice of $\sqrt{Z}_U$, their eigenvalues do not,
showing again that the ambiguity does not affect physical quantities.



\section{\label{sec:gau}Results for the gauge coupling beta functions}

In this section we present the analytical results for the gauge coupling beta
functions of a general 2HDM. We notice that both the Yukawa matrices and the self
couplings occur in the gauge beta functions only through flavour invariants\footnote{
Some of them are identical to the invariants listed in Eq.~\eqref{eq::SMinv}.},
as expected.

We present the results keeping the full information contained in the Yukawa
matrices and arrive at the following expressions for the beta functions

\begin{align}
\beta_{1} &= -\epsilon\frac{\alpha_1}{\pi} + \frac{\alpha_1^2}{(4\pi)^2}\Bigg[\frac{16}{3}n_G + \frac{2}{5}n_D\Bigg]\nonumber\\
                 &+ \frac{\alpha_1^2}{(4\pi)^3}\Bigg[n_G\Big(\frac{76\alpha_1}{15} + \frac{12\alpha_2}{5} + \frac{176\alpha_3}{15}\Big)
                                                   + n_D\Big(\frac{18\alpha_1}{25} + \frac{18\alpha_2}{5}\Big)\nonumber\\
                 &\phantom{+ \frac{\alpha_1^2}{(4\pi)^3}\Bigg[} - \frac{34}{5}\,\mathrm{Tr}(\hat{M}_u)-2\,\mathrm{Tr}{\hat{M}_d}-6\,\mathrm{Tr}{\hat{M}_l}\Bigg]\nonumber\\
                 &+ \frac{\alpha_1^2}{(4\pi)^4}\Bigg[n_G^2\Big(-\frac{836\alpha_1^2}{135} - \frac{44\alpha_2^2}{15} - \frac{1936\alpha_3^2}{135}\Big)
                                                     + n_D^2\Big(-\frac{147\alpha_1^2}{1000} - \frac{49\alpha_2^2}{40}\Big)\nonumber\\
                 &\phantom{+ \frac{\alpha_1^2}{(4\pi)^4}\Bigg[} + n_G n_D \Big(-\frac{887\alpha_1^2}{450} - \frac{173\alpha_2^2}{30} \Big)
                                                                + n_D \Big(\frac{783\alpha_1^2}{2000} + \frac{783\alpha_1\alpha_2}{200} + \frac{3499\alpha_2^2}{80}\Big)\nonumber\\
                 &\phantom{+ \frac{\alpha_1^2}{(4\pi)^4}\Bigg[} + n_G \Big(-\frac{101\alpha_1^2}{90}
                   - \frac{7\alpha_1\alpha_2}{25} - \frac{548\alpha_1\alpha_3}{225} + \frac{101\alpha_2^2}{6} - \frac{4\alpha_2\alpha_3}{5} + \frac{1100\alpha_3^2}{9}\Big)\nonumber\\
                 &\phantom{+ \frac{\alpha_1^2}{(4\pi)^4}\Bigg[} - \frac{2827\alpha_1}{200}\,\mathrm{Tr}(\hat{M}_u) - \frac{1267\alpha_1}{200}\,\mathrm{Tr}(\hat{M}_d) - \frac{2529\alpha_1}{200}\,\mathrm{Tr}(\hat{M}_l)
                                                                - \frac{471\alpha_2}{8}\,\mathrm{Tr}(\hat{M}_u)\nonumber\\
                 &\phantom{+ \frac{\alpha_1^2}{(4\pi)^4}\Bigg[} - \frac{1311\alpha_2}{40}\,\mathrm{Tr}(\hat{M}_d) - \frac{1629\alpha_2}{40}\,\mathrm{Tr}(\hat{M}_l) - \frac{116\alpha_3}{5}\,\mathrm{Tr}(\hat{M}_u)
                                                                - \frac{68\alpha_3}{5}\,\mathrm{Tr}(\hat{M}_d)\nonumber\\
                 &\phantom{+ \frac{\alpha_1^2}{(4\pi)^4}\Bigg[} + \frac{9\alpha_1}{25}\hat{\lambda}_{ij,ji} + \frac{18\alpha_1}{25}\hat{\lambda}_{ii,jj} + \frac{9\alpha_2}{5}\hat{\lambda}_{ij,ji}
                                                                + \frac{213}{20}\,\mathrm{Tr}(\hat{M}_u^2) + \frac{81}{20}\,\mathrm{Tr}(\hat{M}_d^2)\nonumber\\
                 &\phantom{+ \frac{\alpha_1^2}{(4\pi)^4}\Bigg[} + \frac{3}{10}\,\mathrm{Tr}(\hat{M}_u \hat{M}_d)
                                                                + \frac{63}{10}\,\mathrm{Tr}(\hat{M}^{(2)}_{uu}) + \frac{51}{10}\,\mathrm{Tr}(\hat{M}^{(2)}_{dd}) + \frac{6}{5}\,\mathrm{Tr}(\hat{M}^{(2)}_{ud})\nonumber\\
                 &\phantom{+ \frac{\alpha_1^2}{(4\pi)^4}\Bigg[} + \frac{147}{20}\,\mathrm{Tr}(\hat{M}^2_{l}) + \frac{57}{10}\,\mathrm{Tr}(\hat{M}^{(2)}_{ll}) + \frac{303}{10}\hat{T}^{(2)}_{uu} + \frac{51}{10}\hat{T}^{(2)}_{dd}
                                                                + \frac{177}{5}\hat{T}^{(2)}_{ud} + \frac{99}{10}\hat{T}^{(2)}_{ll}\nonumber\\
                 &\phantom{+ \frac{\alpha_1^2}{(4\pi)^4}\Bigg[} + \frac{199}{5}\hat{T}^{(2)}_{ul} + \frac{157}{5}\hat{T}^{(2)}_{dl} - \frac{6}{5}\hat{\lambda}_{ij,kl}\hat{\lambda}_{ji,lk}
                                                                - \frac{3}{5}\hat{\lambda}_{ij,kl}\hat{\lambda}_{li,jk}\Bigg]
\,,
\end{align}
\begin{align}
\beta_{2} &= -\epsilon\frac{\alpha_2}{\pi} + \frac{\alpha_2^2}{(4\pi)^2}\Bigg[-\frac{88}{3} + \frac{16}{3}n_G + \frac{2}{3}n_D\Bigg]\nonumber\\
                 &+ \frac{\alpha_2^2}{(4\pi)^3}\Bigg[-\frac{544\alpha_2}{3} + n_G\Big(\frac{4\alpha_1}{5} + \frac{196\alpha_2}{3} + 16\alpha_3\Big)
                                                   + n_D\Big(\frac{6\alpha_1}{5} + \frac{26\alpha_2}{3}\Big)\nonumber\\
                 &\phantom{+ \frac{\alpha_2^2}{(4\pi)^3}\Bigg[} - 6\,\mathrm{Tr}(\hat{M}_u + \hat{M}_d)-2\,\mathrm{Tr}{\hat{M}_l}\Bigg]\nonumber\\
                 &+ \frac{\alpha_2^2}{(4\pi)^4}\Bigg[-\frac{45712\alpha_2^2}{27} + n_G^2\Big(-\frac{44\alpha_1^2}{45} - \frac{1660\alpha_2^2}{27} - \frac{176\alpha_3^2}{9}\Big)
                                                     + n_D^2\Big(-\frac{49\alpha_1^2}{200} - \frac{425\alpha_2^2}{216}\Big)\nonumber\\
                 &\phantom{+ \frac{\alpha_2^2}{(4\pi)^4}\Bigg[} + n_G n_D \Big(-\frac{91\alpha_1^2}{50} - \frac{1121\alpha_2^2}{54} \Big)
                                                                + n_D \Big(\frac{261\alpha_1^2}{400} + \frac{561\alpha_1\alpha_2}{40} + \frac{65131\alpha_2^2}{432}\Big)\nonumber\\
                 &\phantom{+ \frac{\alpha_2^2}{(4\pi)^4}\Bigg[} + n_G \Big(-\frac{7\alpha_1^2}{150}
                   + \frac{13\alpha_1\alpha_2}{5} - \frac{4\alpha_1\alpha_3}{15} + \frac{52417\alpha_2^2}{54} + 52\alpha_2\alpha_3 + \frac{500\alpha_3^2}{3}\Big)\nonumber\\
                 &\phantom{+ \frac{\alpha_2^2}{(4\pi)^4}\Bigg[} - \frac{593\alpha_1}{40}\,\mathrm{Tr}(\hat{M}_u) - \frac{533\alpha_1}{40}\,\mathrm{Tr}(\hat{M}_d) - \frac{51\alpha_1}{8}\,\mathrm{Tr}(\hat{M}_l)\nonumber\\
                 &\phantom{+ \frac{\alpha_2^2}{(4\pi)^4}\Bigg[}
                                                                - \frac{729\alpha_2}{8}\,\mathrm{Tr}(\hat{M}_u + \hat{M}_d) - \frac{243\alpha_2}{8}\,\mathrm{Tr}(\hat{M}_l) - 28\alpha_3\,\mathrm{Tr}(\hat{M}_u + \hat{M}_d)\nonumber\\
                 &\phantom{+ \frac{\alpha_2^2}{(4\pi)^4}\Bigg[} + \frac{3\alpha_1}{5}\hat{\lambda}_{ij,ji} + \alpha_2\hat{\lambda}_{ij,ji} + 2\alpha_2\hat{\lambda}_{ii,jj} + \frac{15}{4}\,\mathrm{Tr}((\hat{M}_u + \hat{M}_d)^2)\nonumber\\
                 &\phantom{+ \frac{\alpha_2^2}{(4\pi)^4}\Bigg[} + \frac{21}{2}\,\mathrm{Tr}(\hat{M}^{(2)}_{uu} + \hat{M}^{(2)}_{dd}) + 6\,\mathrm{Tr}(\hat{M}^{(2)}_{ud})
                                                                + \frac{5}{4}\,\mathrm{Tr}(\hat{M}^2_{l}) + \frac{7}{2}\,\mathrm{Tr}(\hat{M}^{(2)}_{ll})\nonumber\\
                 &\phantom{+ \frac{\alpha_2^2}{(4\pi)^4}\Bigg[} + \frac{45}{2}(\hat{T}^{(2)}_{uu} + \hat{T}^{(2)}_{dd} + 2\hat{T}^{(2)}_{ud}) + \frac{5}{2}\hat{T}^{(2)}_{ll} + 15(\hat{T}^{(2)}_{ul} + \hat{T}^{(2)}_{dl})\nonumber\\
                 &\phantom{+ \frac{\alpha_2^2}{(4\pi)^4}\Bigg[}
                           - 2\hat{\lambda}_{ij,kl}\hat{\lambda}_{ji,lk} - \hat{\lambda}_{ij,kl}\hat{\lambda}_{li,jk}\Bigg]
\,,
\end{align}
and
\begin{align}
\beta_{3} &= -\epsilon\frac{\alpha_3}{\pi} + \frac{\alpha_3^2}{(4\pi)^2}\Bigg[-44 + \frac{16}{3}n_G \Bigg]\nonumber\\
                 &+ \frac{\alpha_3^2}{(4\pi)^3}\Bigg[-408\alpha_3 + n_G\Big(\frac{22\alpha_1}{15} + 6\alpha_2 + \frac{304\alpha_3}{3}\Big) - 8\,\mathrm{Tr}(\hat{M}_u + \hat{M}_d)\Bigg]\nonumber\\
                 &+ \frac{\alpha_3^2}{(4\pi)^4}\Bigg[-5714\alpha_3^2 + n_G^2\Big(-\frac{242\alpha_1^2}{135} - \frac{22\alpha_2^2}{3} - \frac{2600\alpha_3^2}{27}\Big)
                                                     + n_G n_D \Big(-\frac{253\alpha_1^2}{900} - \frac{23\alpha_2^2}{12} \Big)\nonumber\\
                 &\phantom{+ \frac{\alpha_3^2}{(4\pi)^4}\Bigg[} + n_G \Big(-\frac{137\alpha_1^2}{900}
                   - \frac{\alpha_1\alpha_2}{10} + \frac{308\alpha_1\alpha_3}{45} + \frac{505\alpha_2^2}{12} + 28\alpha_2\alpha_3 + \frac{20132\alpha_3^2}{9}\Big)\nonumber\\
                 &\phantom{+ \frac{\alpha_3^2}{(4\pi)^4}\Bigg[} - \frac{101\alpha_1}{10}\,\mathrm{Tr}(\hat{M}_u) - \frac{89\alpha_1}{10}\,\mathrm{Tr}(\hat{M}_d) - \frac{93\alpha_2}{2}\,\mathrm{Tr}(\hat{M}_u + \hat{M}_d)
                                                                \nonumber\\
                 &\phantom{+ \frac{\alpha_3^2}{(4\pi)^4}\Bigg[} - 160\alpha_3\,\mathrm{Tr}(\hat{M}_u + \hat{M}_d) + 6\,\mathrm{Tr}((\hat{M}_u + \hat{M}_d)^2) + 12\,\mathrm{Tr}(\hat{M}^{(2)}_{uu} + \hat{M}^{(2)}_{dd} - 2\hat{M}^{(2)}_{ud})\nonumber\\
                 &\phantom{+ \frac{\alpha_3^2}{(4\pi)^4}\Bigg[} + 42(\hat{T}^{(2)}_{uu} + \hat{T}^{(2)}_{dd} + 2\hat{T}^{(2)}_{ud}) + 14(\hat{T}^{(2)}_{ul} + \hat{T}^{(2)}_{dl})\Bigg]
\,.
\end{align}

In the above equations $n_G$ denotes the number of fermion generations and
$n_D$ the number of scalar doublets.  We sum over the indices $i,j,k,l$ of the
quartic couplings from 1 to $n_D$. 
The matrix $M_l$ is defined by
\begin{align}
M_l = Y^l_1Y^{l \dagger}_1+Y^l_2Y^{l \dagger}_2
\end{align}
in analogy to Eq.~\eqref{eq::MuMd}. Other combinations of Yukawa matrices are given by
\begin{align}
M^{(2)}_{uu} &= \sum_{i,j=1}^{n_D} Y^u_iY^{u \dagger}_j Y^u_jY^{u \dagger}_i~,\nonumber\\
M^{(2)}_{dd} &= \sum_{i,j=1}^{n_D} Y^d_iY^{d \dagger}_j Y^d_jY^{d \dagger}_i~,\nonumber\\
M^{(2)}_{ll} &= \sum_{i,j=1}^{n_D} Y^l_iY^{l \dagger}_j Y^l_jY^{l \dagger}_i~,\nonumber\\
M^{(2)}_{ud} &= \sum_{i,j=1}^{n_D} Y^u_iY^{u \dagger}_j Y^d_iY^{d \dagger}_j~,
\label{eq::m2}
\end{align}
as well as
\begin{align}
T^{(2)}_{ll} &= \sum_{i,j=1}^{n_D} \mathrm{Tr}(Y^l_iY^{l \dagger}_j)\mathrm{Tr}(Y^l_jY^{l \dagger}_i)~,\nonumber\\
T^{(2)}_{ud} &= \sum_{i,j=1}^{n_D} \mathrm{Tr}(Y^u_iY^{u \dagger}_j)\mathrm{Tr}(Y^d_iY^{d \dagger}_j)~,\nonumber\\
T^{(2)}_{ul} &= \sum_{i,j=1}^{n_D} \mathrm{Tr}(Y^u_iY^{u \dagger}_j)\mathrm{Tr}(Y^l_iY^{l \dagger}_j)~,\nonumber\\
T^{(2)}_{dl} &= \sum_{i,j=1}^{n_D} \mathrm{Tr}(Y^d_iY^{d \dagger}_j)\mathrm{Tr}(Y^l_jY^{l \dagger}_i)~.
\label{eq::t2}
\end{align}
which are defined in analogy to Eqs.~\eqref{eq:T2uu} and \eqref{eq:T2dd}.
We rescaled the Yukawa matrices in the above results such that $\hat{M}_f = M_f/(4\pi)$,
$\hat{T}^{(2)}_{ff'} = T^{(2)}_{ff'}/(4\pi)^2$ and $\hat{M}^{(2)}_{ff'} = M^{(2)}_{ff'}/(4\pi)^2$.
The results for the beta functions and the corresponding renormalization
constants can be obtained in computer readable
form~\cite{progdata}.

We have performed a number of cross checks on the correctness of our result.
Among them is the independence on the three gauge parameters.  Furthermore, we
can easily take the SM limit by setting $n_D = 1$,~$Y^f_2 =
0$ and $\lambda_{ij,kl} = \lambda$ and find agreement with
Refs.~\cite{Mihaila:2012fm,Mihaila:2012pz,Bednyakov:2012rb}.  We also agree
with the findings of Ref.~\cite{Pickering:2001aq} where results for a general
theory based on a simple gauge group are presented.\footnote{In this context
  one has to take into account the comments presented at the end of Section~IV
  of Ref.~\cite{Mihaila:2012pz}.}

A comment on the validity of our results for $n_D \geq 3$ is in order.
At three-loop order, all diagrams containing at least one internal gauge boson or a closed fermion
loop can only receive contributions from up to two different scalar doublets.
However, diagrams containing two quartic couplings can get
contributions from more than two doublets.  Therefore, all contributions to
the three-loop beta functions are also valid for $n_D \geq 3$ apart from those
containing two quartic couplings. 


\section{\label{sec:yuk}Results for the Yukawa coupling beta functions}

As discussed in Section~\ref{sec:ren} the Yukawa matrix beta functions
themselves are ambiguous and one should either work in a proper basis or only
consider invariants of the Yukawa sector.  In general the expressions are
quite lengthy at three loops. Thus, we restrict ourselves to the beta function
in the simplified model discussed in Section~\ref{sec:simplified}.  For
simplicity we drop the primes introduced in Eq.~(\ref{eq::ytprime}) and write
$y^t \equiv y_1^t$ since $\beta_{y_2^t}=0$.  We obtain

{\scalefont{0.8}
\begin{align}
\beta_{y^t} &= -\epsilon\frac{y^t}{2} + \frac{y^t}{4\pi}\Bigg[-\frac{17\alpha_1}{40} - \frac{9\alpha_2}{8} - 4\alpha_3 + \frac{9{\hat{y}^t}{\vphantom{y^t}}^2}{4} + \frac{3\left|{\hat{y}^b_1}\right|^2}{4} 
                                                              +\frac{\left|{\hat{y}^b_2}\right|^2}{4}\Bigg]\nonumber\\
  &+ \frac{y^t}{(4\pi)^2}\Bigg[\frac{107\alpha_1^2}{1200} - \frac{9\alpha_1\alpha_2}{40} + \frac{19\alpha_1\alpha_3}{30} - \frac{33\alpha_2^2}{8}
                             + \frac{9\alpha_2\alpha_3}{2} - \frac{202\alpha_3^2}{3} + n_G\Big(\frac{29\alpha_1^2}{90} + \frac{\alpha_2^2}{2}\nonumber\\
  &\phantom{+ \frac{y^t}{(4\pi)^2}\Bigg[}+ \frac{40\alpha_3^2}{9}\Big)
                                         + \alpha_1 \Big(\frac{393{\hat{y}^t}{\vphantom{y^t}}^2}{160} + \frac{7\left|{\hat{y}^b_1}\right|^2}{160} - \frac{41\left|{\hat{y}^b_2}\right|^2}{480} \Big)
                                         + \alpha_2 \Big(\frac{225{\hat{y}^t}{\vphantom{y^t}}^2}{32} + \frac{99\left|{\hat{y}^b_1}\right|^2}{32}\nonumber\\
  &\phantom{+ \frac{y^t}{(4\pi)^2}\Bigg[}+ \frac{33\left|{\hat{y}^b_2}\right|^2}{32} \Big)
                                         + \alpha_3 \Big(18{\hat{y}^t}{\vphantom{y^t}}^2 + 2\left|{\hat{y}^b_1}\right|^2 + \frac{8\left|{\hat{y}^b_2}\right|^2}{3} \Big)- 6{\hat{y}^t}{\vphantom{y^t}}^4
                                         - \frac{11{\hat{y}^t}{\vphantom{y^t}}^2\left|{\hat{y}^b_1}\right|^2}{8} - \frac{5{\hat{y}^t}{\vphantom{y^t}}^2\left|{\hat{y}^b_2}\right|^2}{4}\nonumber\\
  &\phantom{+ \frac{y^t}{(4\pi)^2}\Bigg[}- \frac{\left|{\hat{y}^b_1}\right|^4}{8}
                                         - \frac{11\left|{\hat{y}^b_1}\right|^2\left|{\hat{y}^b_2}\right|^2}{8} - \frac{5\left|{\hat{y}^b_2}\right|^4}{4} + 3\hat{\lambda}_{11,11}^2
                                         + \frac{\hat{\lambda}_{11,22}^2}{2} + \frac{\hat{\lambda}_{12,21}^2}{2} + \frac{\hat{\lambda}_{11,22}\hat{\lambda}_{12,21}}{2}\nonumber\\
  &\phantom{+ \frac{y^t}{(4\pi)^2}\Bigg[}+ 3\hat{\lambda}_{12,12}\hat{\lambda}_{21,21} + \frac{9\hat{\lambda}_{11,12}\hat{\lambda}_{11,21}}{4}
                                         + \frac{3\hat{\lambda}_{12,22}\hat{\lambda}_{21,22}}{4} - 6{\hat{y}^t}{\vphantom{y^t}}^2\hat{\lambda}_{11,11}
                                         - \left|{\hat{y}^b_2}\right|^2\hat{\lambda}_{11,22}\nonumber\\
  &\phantom{+ \frac{y^t}{(4\pi)^2}\Bigg[}+ \left|{\hat{y}^b_2}\right|^2\hat{\lambda}_{12,21}\Bigg]\nonumber\\
  &+ \frac{y^t}{(4\pi)^3}\Bigg[\frac{3701\alpha_1^3}{6000} + \frac{777\alpha_1^2\alpha_2}{400} - \frac{859\alpha_1^2\alpha_3}{400}
                             + \frac{687\alpha_1\alpha_2^2}{320} - \frac{321\alpha_1\alpha_2\alpha_3}{40} - \frac{127\alpha_1\alpha_3^2}{120}\nonumber\\
  &\phantom{+ \frac{y^t}{(4\pi)^2}\Bigg[}- \frac{699\alpha_2^3}{64} + \frac{501\alpha_2^2\alpha_3}{16} + \frac{531\alpha_2\alpha_3^2}{8}
                                         - 1249\alpha_3^3 + \zeta_3\Big(-\frac{153\alpha_1^3}{1000} - \frac{153\alpha_1^2\alpha_2}{200} \nonumber\\
  &\phantom{+ \frac{y^t}{(4\pi)^2}\Bigg[}- \frac{27\alpha_1\alpha_2^2}{40}+ \frac{45\alpha_2^3}{8}\Big)
                                         + n_G\Big(\frac{56861\alpha_1^3}{21600} + \frac{241\alpha_1^2\alpha_2}{800} + \frac{5281\alpha_1^2\alpha_3}{1800} - \frac{9\alpha_1\alpha_2^2}{160}\nonumber\\
  &\phantom{+ \frac{y^t}{(4\pi)^2}\Bigg[}+ \frac{44\alpha_1\alpha_3^2}{9} - \frac{99\alpha_2^3}{32} + \frac{57\alpha_2^2\alpha_3}{8} + 19\alpha_2\alpha_3^2 + \frac{4432\alpha_3^3}{27}\Big)
                                         + n_G\zeta_3\Big(-\frac{323\alpha_1^3}{150}\nonumber\\
  &\phantom{+ \frac{y^t}{(4\pi)^2}\Bigg[}- \frac{51\alpha_1^2\alpha_2}{50} - \frac{374\alpha_1^2\alpha_3}{75} - \frac{9\alpha_1\alpha_2^2}{10}
                                         - \frac{88\alpha_1\alpha_3^2}{15} + \frac{45\alpha_2^3}{2} - 18\alpha_2^2\alpha_3 - 24\alpha_2\alpha_3^2\nonumber\\
  &\phantom{+ \frac{y^t}{(4\pi)^2}\Bigg[}+ \frac{320\alpha_3^3}{3}\Big) + n_G^2\Big(\frac{73\alpha_1^3}{81} + \frac{25\alpha_2^3}{9} + \frac{560\alpha_3^3}{81}\Big)
                                         + \alpha_1^2\Big(-\frac{69721{\hat{y}^t}{\vphantom{y^t}}^2}{38400} - \frac{7159\left|{\hat{y}^b_1}\right|^2}{7680} \nonumber\\
  &\phantom{+ \frac{y^t}{(4\pi)^2}\Bigg[}- \frac{101419\left|{\hat{y}^b_2}\right|^2}{115200}
                                         - \frac{1089\hat{\lambda}_{11,11}}{800} - \frac{363\hat{\lambda}_{11,22}}{800} - \frac{363\hat{\lambda}_{12,21}}{1600}
                                         + \zeta_3\Big(-\frac{93{\hat{y}^t}{\vphantom{y^t}}^2}{400} \nonumber\\
  &\phantom{+ \frac{y^t}{(4\pi)^2}\Bigg[}- \frac{199\left|{\hat{y}^b_1}\right|^2}{400} + \frac{11\left|{\hat{y}^b_2}\right|^2}{400}\Big)
                                         + n_G\Big(-\frac{115{\hat{y}^t}{\vphantom{y^t}}^2}{32} - \frac{23\left|{\hat{y}^b_1}\right|^2}{480} - \frac{17\left|{\hat{y}^b_2}\right|^2}{96}\Big) \Big)\nonumber\\
  &\phantom{+ \frac{y^t}{(4\pi)^2}\Bigg[}+ \alpha_1\alpha_2\Big(\frac{8097{\hat{y}^t}{\vphantom{y^t}}^2}{1280} + \frac{747\left|{\hat{y}^b_1}\right|^2}{256} + \frac{773\left|{\hat{y}^b_2}\right|^2}{1280}
                                         + \frac{117\hat{\lambda}_{11,11}}{80} + \frac{117\hat{\lambda}_{12,21}}{160}\nonumber\\
  &\phantom{+ \frac{y^t}{(4\pi)^2}\Bigg[}+ \zeta_3\Big(\frac{369{\hat{y}^t}{\vphantom{y^t}}^2}{40} + \frac{27\left|{\hat{y}^b_1}\right|^2}{20} + \frac{9\left|{\hat{y}^b_2}\right|^2}{20}\Big)\Big)
                                         + \alpha_1\alpha_3\Big(-\frac{63{\hat{y}^t}{\vphantom{y^t}}^2}{5} - \frac{457\left|{\hat{y}^b_1}\right|^2}{60}\nonumber\\
  &\phantom{+ \frac{y^t}{(4\pi)^2}\Bigg[}- \frac{259\left|{\hat{y}^b_2}\right|^2}{72} + \zeta_3\Big(18{\hat{y}^t}{\vphantom{y^t}}^2 - \frac{14\left|{\hat{y}^b_1}\right|^2}{5}
                                         + \frac{16\left|{\hat{y}^b_2}\right|^2}{5}\Big)\Big) + \alpha_2^2\Big(\frac{47649{\hat{y}^t}{\vphantom{y^t}}^2}{512} \nonumber\\
  &\phantom{+ \frac{y^t}{(4\pi)^2}\Bigg[}+ \frac{13155\left|{\hat{y}^b_1}\right|^2}{512} + \frac{4329\left|{\hat{y}^b_2}\right|^2}{512}
                                         - \frac{171\hat{\lambda}_{11,11}}{32} - \frac{57\hat{\lambda}_{11,22}}{32} - \frac{57\hat{\lambda}_{12,21}}{64}\nonumber\\
  &\phantom{+ \frac{y^t}{(4\pi)^2}\Bigg[}+ \zeta_3\Big(-\frac{729{\hat{y}^t}{\vphantom{y^t}}^2}{16} - \frac{225\left|{\hat{y}^b_1}\right|^2}{32} - \frac{99\left|{\hat{y}^b_2}\right|^2}{32}\Big)
                                         + n_G\Big(-\frac{351{\hat{y}^t}{\vphantom{y^t}}^2}{32} - \frac{69\left|{\hat{y}^b_1}\right|^2}{16} \nonumber\\
  &\phantom{+ \frac{y^t}{(4\pi)^2}\Bigg[}- \frac{63\left|{\hat{y}^b_2}\right|^2}{16}\Big) \Big) + \alpha_2\alpha_3\Big(-84{\hat{y}^t}{\vphantom{y^t}}^2 - \frac{27\left|{\hat{y}^b_1}\right|^2}{4} + \frac{37\left|{\hat{y}^b_2}\right|^2}{8}
                                         + \zeta_3\Big(90{\hat{y}^t}{\vphantom{y^t}}^2 - 54\left|{\hat{y}^b_1}\right|^2\Big)\Big)\nonumber\\
  &\phantom{+ \frac{y^t}{(4\pi)^2}\Bigg[}+ \alpha_3^2\Big(\frac{4799{\hat{y}^t}{\vphantom{y^t}}^2}{12} - \frac{277\left|{\hat{y}^b_1}\right|^2}{4} + \frac{2227\left|{\hat{y}^b_2}\right|^2}{36}
                                         + \zeta_3\Big(-114{\hat{y}^t}{\vphantom{y^t}}^2 - 22\left|{\hat{y}^b_1}\right|^2 - 34\left|{\hat{y}^b_2}\right|^2\Big)\nonumber\\
  &\phantom{+ \frac{y^t}{(4\pi)^2}\Bigg[}+ n_G\Big(-27{\hat{y}^t}{\vphantom{y^t}}^2 - \frac{7\left|{\hat{y}^b_1}\right|^2}{3} - \frac{11\left|{\hat{y}^b_2}\right|^2}{3}\Big) \Big)
                                         + \alpha_1\Big(-\frac{2437{\hat{y}^t}{\vphantom{y^t}}^4}{160} - \frac{1383{\hat{y}^t}{\vphantom{y^t}}^2\left|{\hat{y}^b_1}\right|^2}{320} \nonumber\\
  &\phantom{+ \frac{y^t}{(4\pi)^2}\Bigg[}- \frac{697{\hat{y}^t}{\vphantom{y^t}}^2\left|{\hat{y}^b_2}\right|^2}{480}
                                         - \frac{959\left|{\hat{y}^b_1}\right|^4}{320} - \frac{1133\left|{\hat{y}^b_1}\right|^2\left|{\hat{y}^b_2}\right|^2}{960}
                                         + \frac{109\left|{\hat{y}^b_2}\right|^4}{60} + \zeta_3\Big(\frac{{\hat{y}^t}{\vphantom{y^t}}^2\left|{\hat{y}^b_1}\right|^2}{4} \nonumber\\
  &\phantom{+ \frac{y^t}{(4\pi)^2}\Bigg[}+ \frac{3{\hat{y}^t}{\vphantom{y^t}}^2\left|{\hat{y}^b_2}\right|^2}{4}
                                         + \frac{19\left|{\hat{y}^b_1}\right|^4}{20} - \frac{2\left|{\hat{y}^b_1}\right|^2\left|{\hat{y}^b_2}\right|^2}{5} - \frac{27\left|{\hat{y}^b_2}\right|^4}{20}\Big)
                                         - \frac{127{\hat{y}^t}{\vphantom{y^t}}^2\hat{\lambda}_{11,11}}{20}\nonumber\\
  &\phantom{+ \frac{y^t}{(4\pi)^2}\Bigg[}- \frac{139\left|{\hat{y}^b_2}\right|^2\hat{\lambda}_{11,22}}{120} + \frac{139\left|{\hat{y}^b_2}\right|^2\hat{\lambda}_{12,21}}{120}
                                         + \frac{9\hat{\lambda}_{11,11}^2}{2} + \frac{3\hat{\lambda}_{11,22}^2}{4} + \frac{3\hat{\lambda}_{12,21}^2}{4}\nonumber\\
  &\phantom{+ \frac{y^t}{(4\pi)^2}\Bigg[}+ \frac{3\hat{\lambda}_{11,22}\hat{\lambda}_{12,21}}{4} + \frac{9\hat{\lambda}_{12,12}\hat{\lambda}_{21,21}}{2}
                                         + \frac{27\hat{\lambda}_{11,12}\hat{\lambda}_{11,21}}{8} + \frac{9\hat{\lambda}_{12,22}\hat{\lambda}_{21,22}}{8} \Big)\nonumber\\
  &\phantom{+ \frac{y^t}{(4\pi)^2}\Bigg[}+ \alpha_2\Big(-\frac{1593{\hat{y}^t}{\vphantom{y^t}}^4}{32} - \frac{2307{\hat{y}^t}{\vphantom{y^t}}^2\left|{\hat{y}^b_1}\right|^2}{64} - \frac{141{\hat{y}^t}{\vphantom{y^t}}^2\left|{\hat{y}^b_2}\right|^2}{32}
                                         - \frac{2283\left|{\hat{y}^b_1}\right|^4}{64} - \frac{2763\left|{\hat{y}^b_1}\right|^2\left|{\hat{y}^b_2}\right|^2}{64}\nonumber\\
  &\phantom{+ \frac{y^t}{(4\pi)^2}\Bigg[}- \frac{15\left|{\hat{y}^b_2}\right|^4}{2} + \zeta_3\Big(-\frac{9{\hat{y}^t}{\vphantom{y^t}}^2\left|{\hat{y}^b_1}\right|^2}{4} - \frac{9{\hat{y}^t}{\vphantom{y^t}}^2\left|{\hat{y}^b_2}\right|^2}{4}
                                         + \frac{63\left|{\hat{y}^b_1}\right|^4}{4} + 18\left|{\hat{y}^b_1}\right|^2\left|{\hat{y}^b_2}\right|^2 + \frac{9\left|{\hat{y}^b_2}\right|^4}{4}\Big)\nonumber\\
  &\phantom{+ \frac{y^t}{(4\pi)^2}\Bigg[}- \frac{135{\hat{y}^t}{\vphantom{y^t}}^2\hat{\lambda}_{11,11}}{4} - \frac{45\left|{\hat{y}^b_2}\right|^2\hat{\lambda}_{11,22}}{8} + \frac{45\left|{\hat{y}^b_2}\right|^2\hat{\lambda}_{12,21}}{8}
                                         + \frac{45\hat{\lambda}_{11,11}^2}{2} + \frac{15\hat{\lambda}_{11,22}^2}{4} + \frac{15\hat{\lambda}_{12,21}^2}{4}\nonumber\\
  &\phantom{+ \frac{y^t}{(4\pi)^2}\Bigg[}+ \frac{15\hat{\lambda}_{11,22}\hat{\lambda}_{12,21}}{4} + \frac{45\hat{\lambda}_{12,12}\hat{\lambda}_{21,21}}{2}
                                         + \frac{135\hat{\lambda}_{11,12}\hat{\lambda}_{11,21}}{8} + \frac{45\hat{\lambda}_{12,22}\hat{\lambda}_{21,22}}{8} \Big)\nonumber\\
  &\phantom{+ \frac{y^t}{(4\pi)^2}\Bigg[}+ \alpha_3\Big(-\frac{157{\hat{y}^t}{\vphantom{y^t}}^4}{2} + \frac{27{\hat{y}^t}{\vphantom{y^t}}^2\left|{\hat{y}^b_1}\right|^2}{2} - \frac{185{\hat{y}^t}{\vphantom{y^t}}^2\left|{\hat{y}^b_2}\right|^2}{12}
                                         + 41\left|{\hat{y}^b_1}\right|^4 + \frac{235\left|{\hat{y}^b_1}\right|^2\left|{\hat{y}^b_2}\right|^2}{12}\nonumber\\
  &\phantom{+ \frac{y^t}{(4\pi)^2}\Bigg[}- \frac{257\left|{\hat{y}^b_2}\right|^4}{12} + \zeta_3\Big(-16{\hat{y}^t}{\vphantom{y^t}}^2\left|{\hat{y}^b_1}\right|^2 - 12{\hat{y}^t}{\vphantom{y^t}}^2\left|{\hat{y}^b_2}\right|^2
                                         - 32\left|{\hat{y}^b_1}\right|^4 - 20\left|{\hat{y}^b_1}\right|^2\left|{\hat{y}^b_2}\right|^2 + 12\left|{\hat{y}^b_2}\right|^4\Big)\nonumber\\
  &\phantom{+ \frac{y^t}{(4\pi)^2}\Bigg[}+ 8{\hat{y}^t}{\vphantom{y^t}}^2\hat{\lambda}_{11,11} + \frac{4\left|{\hat{y}^b_2}\right|^2\hat{\lambda}_{11,22}}{3} - \frac{4\left|{\hat{y}^b_2}\right|^2\hat{\lambda}_{12,21}}{3}\Big)
                                         + \frac{339{\hat{y}^t}{\vphantom{y^t}}^6}{16} + \frac{739{\hat{y}^t}{\vphantom{y^t}}^4\left|{\hat{y}^b_1}\right|^2}{32} + \frac{49{\hat{y}^t}{\vphantom{y^t}}^4\left|{\hat{y}^b_2}\right|^2}{16}\nonumber\\
  &\phantom{+ \frac{y^t}{(4\pi)^2}\Bigg[}+ \frac{825{\hat{y}^t}{\vphantom{y^t}}^2\left|{\hat{y}^b_1}\right|^4}{16} + \frac{671{\hat{y}^t}{\vphantom{y^t}}^2\left|{\hat{y}^b_1}\right|^2\left|{\hat{y}^b_2}\right|^2}{16} + \frac{87{\hat{y}^t}{\vphantom{y^t}}^2\left|{\hat{y}^b_2}\right|^4}{8}
                                         + \frac{477\left|{\hat{y}^b_1}\right|^6}{32} + \frac{525\left|{\hat{y}^b_1}\right|^4\left|{\hat{y}^b_2}\right|^2}{16} \nonumber\\
  &\phantom{+ \frac{y^t}{(4\pi)^2}\Bigg[}+ \frac{669\left|{\hat{y}^b_1}\right|^2\left|{\hat{y}^b_2}\right|^4}{32} + 3\left|{\hat{y}^b_2}\right|^6
                                         + \zeta_3\Big(\frac{27{\hat{y}^t}{\vphantom{y^t}}^6}{4} - 24{\hat{y}^t}{\vphantom{y^t}}^2\left|{\hat{y}^b_1}\right|^4 - 24{\hat{y}^t}{\vphantom{y^t}}^2\left|{\hat{y}^b_1}\right|^2\left|{\hat{y}^b_2}\right|^2
                                         + \frac{9\left|{\hat{y}^b_1}\right|^6}{4}\nonumber\\
  &\phantom{+ \frac{y^t}{(4\pi)^2}\Bigg[}+ \frac{21\left|{\hat{y}^b_1}\right|^4\left|{\hat{y}^b_2}\right|^2}{4} + \frac{15\left|{\hat{y}^b_1}\right|^2\left|{\hat{y}^b_2}\right|^4}{4} + \frac{3\left|{\hat{y}^b_2}\right|^6}{4}\Big)
                                         + \hat{\lambda}_{11,11}\Big(99{\hat{y}^t}{\vphantom{y^t}}^4 + \frac{93{\hat{y}^t}{\vphantom{y^t}}^2\left|{\hat{y}^b_1}\right|^2}{2}\nonumber\\
  &\phantom{+ \frac{y^t}{(4\pi)^2}\Bigg[}+ \frac{3{\hat{y}^t}{\vphantom{y^t}}^2\left|{\hat{y}^b_2}\right|^2}{2} + \frac{15\left|{\hat{y}^b_1}\right|^4}{2}
                                         + 6\left|{\hat{y}^b_1}\right|^2\left|{\hat{y}^b_2}\right|^2\Big) + \hat{\lambda}_{11,22}\Big(\frac{23{\hat{y}^t}{\vphantom{y^t}}^2\left|{\hat{y}^b_2}\right|^2}{2} 
                                         + 11\left|{\hat{y}^b_1}\right|^2\left|{\hat{y}^b_2}\right|^2 \nonumber\\
  &\phantom{+ \frac{y^t}{(4\pi)^2}\Bigg[}+ \frac{35\left|{\hat{y}^b_2}\right|^4}{4} \Big) + \hat{\lambda}_{12,21}\Big(-\frac{11{\hat{y}^t}{\vphantom{y^t}}^2\left|{\hat{y}^b_2}\right|^2}{2} 
                                         - \frac{7\left|{\hat{y}^b_1}\right|^2\left|{\hat{y}^b_2}\right|^2}{2} - \frac{23\left|{\hat{y}^b_2}\right|^4}{4} \Big)\nonumber\\
  &\phantom{+ \frac{y^t}{(4\pi)^2}\Bigg[}+ \hat{\lambda}_{11,12}\Big(\frac{57{\hat{y}^t}{\vphantom{y^t}}^2\hat{y}^b_1{\hat{y}^b_2}{\vphantom{y^b_2}}^*}{4} 
                                         + \frac{39\left|{\hat{y}^b_1}\right|^2\hat{y}^b_1{\hat{y}^b_2}{\vphantom{y^b_2}}^*}{8} + 3\left|{\hat{y}^b_2}\right|^2\hat{y}^b_1{\hat{y}^b_2}{\vphantom{y^b_2}}^* \Big)
                                         + \hat{\lambda}_{11,21}\Big(\frac{57{\hat{y}^t}{\vphantom{y^t}}^2{\hat{y}^b_1}{\vphantom{y^b_1}}^*\hat{y}^b_2}{4} \nonumber\\
  &\phantom{+ \frac{y^t}{(4\pi)^2}\Bigg[}+ \frac{39\left|{\hat{y}^b_1}\right|^2{\hat{y}^b_1}{\vphantom{y^b_1}}^*\hat{y}^b_2}{8} + 3\left|{\hat{y}^b_2}\right|^2{\hat{y}^b_1}{\vphantom{y^b_1}}^*\hat{y}^b_2 \Big)+ 3\hat{\lambda}_{22,22}\left|{\hat{y}^b_2}\right|^4\
                                         + \frac{9\hat{\lambda}_{12,12}\left(\hat{y}^b_1{\hat{y}^b_2}{\vphantom{y^b_2}}^*\right)^2}{4}\nonumber\\
  &\phantom{+ \frac{y^t}{(4\pi)^2}\Bigg[}+ \frac{9\hat{\lambda}_{21,21}\left({\hat{y}^b_1}{\vphantom{y^b_1}}^*\hat{y}^b_2\right)^2}{4} + \frac{21\hat{\lambda}_{12,22}\left|{\hat{y}^b_2}\right|^2\hat{y}^b_1{\hat{y}^b_2}{\vphantom{y^b_2}}^*}{8}
                                         + \frac{21\hat{\lambda}_{21,22}\left|{\hat{y}^b_2}\right|^2{\hat{y}^b_1}{\vphantom{y^b_1}}^*\hat{y}^b_2}{8} + \hat{\lambda}_{11,11}^2\Big(\frac{15{\hat{y}^t}{\vphantom{y^t}}^2}{8}\nonumber\\
  &\phantom{+ \frac{y^t}{(4\pi)^2}\Bigg[}- \frac{291\left|{\hat{y}^b_1}\right|^2}{8}\Big)
                                         + \hat{\lambda}_{11,22}^2\Big(-\frac{15{\hat{y}^t}{\vphantom{y^t}}^2}{16} - \frac{37\left|{\hat{y}^b_1}\right|^2}{16} - \frac{39\left|{\hat{y}^b_2}\right|^2}{16}\Big)
                                         + \hat{\lambda}_{12,21}^2\Big(-\frac{39{\hat{y}^t}{\vphantom{y^t}}^2}{16}\nonumber\\
  &\phantom{+ \frac{y^t}{(4\pi)^2}\Bigg[}+ \frac{11\left|{\hat{y}^b_1}\right|^2}{16} - \frac{111\left|{\hat{y}^b_2}\right|^2}{16}\Big)
                                         - \frac{33\hat{\lambda}_{22,22}^2\left|{\hat{y}^b_2}\right|^2}{8} + \hat{\lambda}_{12,12}\hat{\lambda}_{21,21}\Big(-\frac{45{\hat{y}^t}{\vphantom{y^t}}^2}{8}
                                         - \frac{15\left|{\hat{y}^b_1}\right|^2}{8}\nonumber\\
  &\phantom{+ \frac{y^t}{(4\pi)^2}\Bigg[}- \frac{357\left|{\hat{y}^b_2}\right|^2}{8}\Big) + \hat{\lambda}_{11,12}\hat{\lambda}_{11,21}\Big(-\frac{69{\hat{y}^t}{\vphantom{y^t}}^2}{32} - \frac{507\left|{\hat{y}^b_1}\right|^2}{32}
                                         - \frac{447\left|{\hat{y}^b_2}\right|^2}{32}\Big)\nonumber\\
  &\phantom{+ \frac{y^t}{(4\pi)^2}\Bigg[}+ \hat{\lambda}_{12,22}\hat{\lambda}_{21,22}\Big(-\frac{99{\hat{y}^t}{\vphantom{y^t}}^2}{32} + \frac{75\left|{\hat{y}^b_1}\right|^2}{32} - \frac{513\left|{\hat{y}^b_2}\right|^2}{32}\Big)
                                         + \hat{\lambda}_{11,22}\hat{\lambda}_{12,21}\Big(-\frac{15{\hat{y}^t}{\vphantom{y^t}}^2}{16}\nonumber\\
  &\phantom{+ \frac{y^t}{(4\pi)^2}\Bigg[}- \frac{37\left|{\hat{y}^b_1}\right|^2}{16} - \frac{135\left|{\hat{y}^b_2}\right|^2}{16}\Big)
                                         + 3\hat{\lambda}_{11,11}\hat{\lambda}_{11,22}\left|{\hat{y}^b_2}\right|^2 + 3\hat{\lambda}_{11,22}\hat{\lambda}_{22,22}\left|{\hat{y}^b_2}\right|^2 \nonumber\\
  &\phantom{+ \frac{y^t}{(4\pi)^2}\Bigg[}- \frac{345\hat{\lambda}_{11,11}\hat{\lambda}_{11,12}\hat{y}^b_1{\hat{y}^b_2}{\vphantom{y^b_2}}^*}{16} - \frac{345\hat{\lambda}_{11,11}\hat{\lambda}_{11,21}{\hat{y}^b_1}{\vphantom{y^b_1}}^*\hat{y}^b_2}{16}
                                         - \frac{159\hat{\lambda}_{11,22}\hat{\lambda}_{11,12}\hat{y}^b_1{\hat{y}^b_2}{\vphantom{y^b_2}}^*}{32}  \nonumber\\
  &\phantom{+ \frac{y^t}{(4\pi)^2}\Bigg[}- \frac{159\hat{\lambda}_{11,22}\hat{\lambda}_{11,21}{\hat{y}^b_1}{\vphantom{y^b_1}}^*\hat{y}^b_2}{32} - \frac{165\hat{\lambda}_{11,22}\hat{\lambda}_{12,22}\hat{y}^b_1{\hat{y}^b_2}{\vphantom{y^b_2}}^*}{32}
                                         - \frac{165\hat{\lambda}_{11,22}\hat{\lambda}_{21,22}{\hat{y}^b_1}{\vphantom{y^b_1}}^*\hat{y}^b_2}{32} \nonumber\\
  &\phantom{+ \frac{y^t}{(4\pi)^2}\Bigg[}- \frac{255\hat{\lambda}_{12,21}\hat{\lambda}_{11,12}\hat{y}^b_1{\hat{y}^b_2}{\vphantom{y^b_2}}^*}{32} - \frac{255\hat{\lambda}_{12,21}\hat{\lambda}_{11,21}{\hat{y}^b_1}{\vphantom{y^b_1}}^*\hat{y}^b_2}{32} 
                                         - \frac{69\hat{\lambda}_{12,21}\hat{\lambda}_{12,22}\hat{y}^b_1{\hat{y}^b_2}{\vphantom{y^b_2}}^*}{32}\nonumber\\
  &\phantom{+ \frac{y^t}{(4\pi)^2}\Bigg[}- \frac{69\hat{\lambda}_{12,21}\hat{\lambda}_{21,22}{\hat{y}^b_1}{\vphantom{y^b_1}}^*\hat{y}^b_2}{32}
                                         + \frac{21\hat{\lambda}_{22,22}\hat{\lambda}_{12,22}\hat{y}^b_1{\hat{y}^b_2}{\vphantom{y^b_2}}^*}{16} + \frac{21\hat{\lambda}_{22,22}\hat{\lambda}_{21,22}{\hat{y}^b_1}{\vphantom{y^b_1}}^*\hat{y}^b_2}{16} \nonumber\\
  &\phantom{+ \frac{y^t}{(4\pi)^2}\Bigg[}- \frac{255\hat{\lambda}_{12,12}\hat{\lambda}_{11,21}\hat{y}^b_1{\hat{y}^b_2}{\vphantom{y^b_2}}^*}{16} - \frac{255\hat{\lambda}_{21,21}\hat{\lambda}_{11,12}{\hat{y}^b_1}{\vphantom{y^b_1}}^*\hat{y}^b_2}{16} \nonumber\\
  &\phantom{+ \frac{y^t}{(4\pi)^2}\Bigg[}- \frac{69\hat{\lambda}_{12,12}\hat{\lambda}_{21,22}\hat{y}^b_1{\hat{y}^b_2}{\vphantom{y^b_2}}^*}{16} - \frac{69\hat{\lambda}_{21,21}\hat{\lambda}_{12,22}{\hat{y}^b_1}{\vphantom{y^b_1}}^*\hat{y}^b_2}{16} 
                                         + \frac{3\hat{\lambda}_{11,12}\hat{\lambda}_{21,22}\left|{\hat{y}^b_2}\right|^2}{2}\nonumber\\
  &\phantom{+ \frac{y^t}{(4\pi)^2}\Bigg[}+ \frac{3\hat{\lambda}_{11,21}\hat{\lambda}_{12,22}\left|{\hat{y}^b_2}\right|^2}{2}- 18\hat{\lambda}_{11,11}^3 - \frac{\hat{\lambda}_{11,22}^3}{2}
                                         - \frac{5\hat{\lambda}_{12,21}^3}{8} - 3\hat{\lambda}_{11,11}\hat{\lambda}_{11,22}^2
                                         - \frac{3\hat{\lambda}_{11,11}\hat{\lambda}_{12,21}^2}{2}\nonumber\\
  &\phantom{+ \frac{y^t}{(4\pi)^2}\Bigg[}- 3\hat{\lambda}_{11,11}\hat{\lambda}_{11,22}\hat{\lambda}_{12,21}- 6\hat{\lambda}_{11,11}\hat{\lambda}_{12,12}\hat{\lambda}_{21,21}
                                         - \frac{45\hat{\lambda}_{11,11}\hat{\lambda}_{11,12}\hat{\lambda}_{11,21}}{2} - \frac{3\hat{\lambda}_{11,22}^2\hat{\lambda}_{22,22}}{2}\nonumber\\
  &\phantom{+ \frac{y^t}{(4\pi)^2}\Bigg[}- \frac{3\hat{\lambda}_{11,22}^2\hat{\lambda}_{12,21}}{4}- \frac{3\hat{\lambda}_{11,22}\hat{\lambda}_{12,21}\hat{\lambda}_{22,22}}{2}
                                         - \frac{3\hat{\lambda}_{11,22}\hat{\lambda}_{12,21}^2}{2} - \frac{9\hat{\lambda}_{11,22}\hat{\lambda}_{11,12}\hat{\lambda}_{11,21}}{2}\nonumber\\
  &\phantom{+ \frac{y^t}{(4\pi)^2}\Bigg[}- \frac{27\hat{\lambda}_{11,22}\hat{\lambda}_{11,12}\hat{\lambda}_{21,22}}{8}- \frac{27\hat{\lambda}_{11,22}\hat{\lambda}_{11,21}\hat{\lambda}_{12,22}}{8}
                                         - 9\hat{\lambda}_{11,22}\hat{\lambda}_{12,12}\hat{\lambda}_{21,21}
                                         - \frac{9\hat{\lambda}_{11,22}\hat{\lambda}_{12,22}\hat{\lambda}_{21,22}}{4}\nonumber\\
  &\phantom{+ \frac{y^t}{(4\pi)^2}\Bigg[}- \frac{3\hat{\lambda}_{12,21}^2\hat{\lambda}_{22,22}}{4}- 6\hat{\lambda}_{12,21}\hat{\lambda}_{11,21}\hat{\lambda}_{11,12}
                                         - \frac{9\hat{\lambda}_{12,21}\hat{\lambda}_{11,12}\hat{\lambda}_{21,22}}{4}- \frac{9\hat{\lambda}_{12,21}\hat{\lambda}_{11,21}\hat{\lambda}_{12,22}}{4}\nonumber\\       
  &\phantom{+ \frac{y^t}{(4\pi)^2}\Bigg[}- \frac{27\hat{\lambda}_{12,21}\hat{\lambda}_{12,12}\hat{\lambda}_{21,21}}{2}- 3\hat{\lambda}_{12,21}\hat{\lambda}_{12,22}\hat{\lambda}_{21,22}
                                         - 3\hat{\lambda}_{22,22}\hat{\lambda}_{12,12}\hat{\lambda}_{21,21}- \frac{15\hat{\lambda}_{11,12}^2\hat{\lambda}_{21,21}}{2}\nonumber\\       
  &\phantom{+ \frac{y^t}{(4\pi)^2}\Bigg[}- \frac{9\hat{\lambda}_{11,12}\hat{\lambda}_{21,21}\hat{\lambda}_{12,22}}{4}- \frac{15\hat{\lambda}_{11,21}^2\hat{\lambda}_{12,12}}{2}
                                         - \frac{9\hat{\lambda}_{11,21}\hat{\lambda}_{12,12}\hat{\lambda}_{21,22}}{4}- \frac{15\hat{\lambda}_{12,12}\hat{\lambda}_{21,22}^2}{4}\nonumber\\       
  &\phantom{+ \frac{y^t}{(4\pi)^2}\Bigg[}- \frac{15\hat{\lambda}_{21,21}\hat{\lambda}_{12,22}^2}{4} - \frac{9\hat{\lambda}_{22,22}\hat{\lambda}_{12,22}\hat{\lambda}_{21,22}}{2} \Bigg]
\,,
\label{eq::beta_yt}
\end{align}
}

\noindent
where also the quartic couplings are given in the new basis (cf. Eq.~\eqref{eq::trafo3}).
We rescaled all Yukawa couplings by $\hat{y} = y/\sqrt{4\pi}$ but the leading ones.
The analytic expression of Eq.~(\ref{eq::beta_yt})
and the beta functions for $y^b_1$ and $y^b_2$ are contained in
ancillary files which come together with this paper~\cite{progdata}.
They also contain the beta functions for the eleven invariants
specified to the 2HDM models I, II, X and Y and the SM.
We furthermore provide explicit results for the quantities in
Eqs.~(\ref{eq::propren}) and~(\ref{eq::ZfYuk}).

Let us mention that for the SM Yukawa matrix beta functions we find full
agreement with~\cite{Bednyakov:2012en,Bednyakov:2014pia} and the one- and
two-loop beta functions for $\mathbb{Z}_2$-symmetric 2HDMs agree
with~\cite{Chowdhury:2015yja}.


\section{\label{sec:sum}Summary}

We consider a general 2HDM and compute the beta functions for the gauge and
Yukawa couplings up to three loops. We discuss in detail the subtleties in
connection to the determination of the renormalization constants in case both
Higgs doublets couple to up- and down-type fermions. Furthermore, we
investigate in detail the origin of the poles in the Yukawa coupling beta
functions, a characteristic which is already present in the SM, discuss their
ambiguity, and provide possible solutions which lead to finite beta functions.
Our general results can be specified to $\mathbb{Z}_2$ symmetric models like
the 2HDMs of type I, II, X or Y, or the SM.  In this paper we also provide the
first independent cross check of the three-loop corrections to the SM Yukawa
coupling beta functions~\cite{Bednyakov:2012en,Bednyakov:2014pia}.  Ancillary
files with analytic results for both, renormalization constants and beta
functions, can be downloaded from~\cite{progdata}.


\section*{Acknowledgements}

This work was supported by Deutsche Forschungsgemeinschaft (contract MI 1358,
Heisenberg program).  F.H. acknowledges the support by the DFG-funded Doctoral
School KSETA.  F.H. would like to thank Thomas Deppisch for fruitful
discussions about numerically solving the RGE for Yukawa matrices.



\begin{thebibliography}{99}
%
%

\bibitem{Mihaila:2012fm}
  L.~N.~Mihaila, J.~Salomon and M.~Steinhauser,
  Phys.\ Rev.\ Lett.\  {\bf 108} (2012) 151602
  doi:10.1103/PhysRevLett.108.151602
  [arXiv:1201.5868 [hep-ph]].

\bibitem{Mihaila:2012pz}
  L.~N.~Mihaila, J.~Salomon and M.~Steinhauser,
  Phys.\ Rev.\ D {\bf 86} (2012) 096008
  doi:10.1103/PhysRevD.86.096008
  [arXiv:1208.3357 [hep-ph]].

\bibitem{Bednyakov:2012rb}
  A.~V.~Bednyakov, A.~F.~Pikelner and V.~N.~Velizhanin,
  JHEP {\bf 1301} (2013) 017
  doi:10.1007/JHEP01(2013)017
  [arXiv:1210.6873 [hep-ph]].

\bibitem{Bednyakov:2012en}
  A.~V.~Bednyakov, A.~F.~Pikelner and V.~N.~Velizhanin,
  Phys.\ Lett.\ B {\bf 722} (2013) 336
  doi:10.1016/j.physletb.2013.04.038
  [arXiv:1212.6829 [hep-ph]].

\bibitem{Bednyakov:2013cpa}
  A.~V.~Bednyakov, A.~F.~Pikelner and V.~N.~Velizhanin,
  Nucl.\ Phys.\ B {\bf 879} (2014) 256
  doi:10.1016/j.nuclphysb.2013.12.012
  [arXiv:1310.3806 [hep-ph]].

\bibitem{Bednyakov:2014pia}
  A.~V.~Bednyakov, A.~F.~Pikelner and V.~N.~Velizhanin,
  Phys.\ Lett.\ B {\bf 737} (2014) 129
  doi:10.1016/j.physletb.2014.08.049
  [arXiv:1406.7171 [hep-ph]].

\bibitem{Chetyrkin:2013wya}
  K.~G.~Chetyrkin and M.~F.~Zoller,
  JHEP {\bf 1304} (2013) 091
   Erratum: [JHEP {\bf 1309} (2013) 155]
  doi:10.1007/JHEP04(2013)091, 10.1007/JHEP09(2013)155
  [arXiv:1303.2890 [hep-ph]].

\bibitem{Bednyakov:2013eba}
  A.~V.~Bednyakov, A.~F.~Pikelner and V.~N.~Velizhanin,
  Nucl.\ Phys.\ B {\bf 875} (2013) 552
  doi:10.1016/j.nuclphysb.2013.07.015
  [arXiv:1303.4364 [hep-ph]].

\bibitem{Martin:2015eia}
  S.~P.~Martin,
  Phys.\ Rev.\ D {\bf 92} (2015) no.5,  054029
  doi:10.1103/PhysRevD.92.054029
  [arXiv:1508.00912 [hep-ph]].

\bibitem{Bednyakov:2015ooa}
  A.~V.~Bednyakov and A.~F.~Pikelner,
  Phys.\ Lett.\ B {\bf 762} (2016) 151
  doi:10.1016/j.physletb.2016.09.007
  [arXiv:1508.02680 [hep-ph]].

\bibitem{Zoller:2015tha}
  M.~F.~Zoller,
  JHEP {\bf 1602} (2016) 095
  doi:10.1007/JHEP02(2016)095
  [arXiv:1508.03624 [hep-ph]].

\bibitem{Chetyrkin:2016ruf}
  K.~G.~Chetyrkin and M.~F.~Zoller,
  JHEP {\bf 1606} (2016) 175
  doi:10.1007/JHEP06(2016)175
  [arXiv:1604.00853 [hep-ph]].

\bibitem{Baikov:2016tgj}
  P.~A.~Baikov, K.~G.~Chetyrkin and J.~H.~Kühn,
  Phys.\ Rev.\ Lett.\  {\bf 118} (2017) no.8,  082002
  doi:10.1103/PhysRevLett.118.082002
  [arXiv:1606.08659 [hep-ph]].

\bibitem{Herzog:2017ohr}
  F.~Herzog, B.~Ruijl, T.~Ueda, J.~A.~M.~Vermaseren and A.~Vogt,
  JHEP {\bf 1702} (2017) 090
  doi:10.1007/JHEP02(2017)090
  [arXiv:1701.01404 [hep-ph]].

\bibitem{Luthe:2017ttg}
  T.~Luthe, A.~Maier, P.~Marquard and Y.~Schroder,
  JHEP {\bf 1710} (2017) 166
  doi:10.1007/JHEP10(2017)166
  [arXiv:1709.07718 [hep-ph]].

\bibitem{Machacek:1983tz}
  M.~E.~Machacek and M.~T.~Vaughn,
  Nucl.\ Phys.\ B {\bf 222} (1983) 83.
  doi:10.1016/0550-3213(83)90610-7

\bibitem{Machacek:1983fi}
  M.~E.~Machacek and M.~T.~Vaughn,
  Nucl.\ Phys.\ B {\bf 236} (1984) 221.
  doi:10.1016/0550-3213(84)90533-9

\bibitem{Machacek:1984zw}
  M.~E.~Machacek and M.~T.~Vaughn,
  Nucl.\ Phys.\ B {\bf 249} (1985) 70.
  doi:10.1016/0550-3213(85)90040-9

\bibitem{Pickering:2001aq}
  A.~G.~M.~Pickering, J.~A.~Gracey and D.~R.~T.~Jones,
  Phys.\ Lett.\ B {\bf 510} (2001) 347
   Erratum: [Phys.\ Lett.\ B {\bf 535} (2002) 377]
  doi:10.1016/S0370-2693(02)01779-3, 10.1016/S0370-2693(01)00624-4
  [hep-ph/0104247].

\bibitem{Branco:2011iw}
  G.~C.~Branco, P.~M.~Ferreira, L.~Lavoura, M.~N.~Rebelo, M.~Sher and
  J.~P.~Silva,
  Phys.\ Rept.\  {\bf 516} (2012) 1
  doi:10.1016/j.physrep.2012.02.002
  [arXiv:1106.0034 [hep-ph]].

\bibitem{Wu:1994ja}
  Y.~L.~Wu and L.~Wolfenstein,
  Phys.\ Rev.\ Lett.\  {\bf 73} (1994) 1762
  doi:10.1103/PhysRevLett.73.1762
  [hep-ph/9409421].

\bibitem{Paschos:1976ay}
  E.~A.~Paschos,
  Phys.\ Rev.\ D {\bf 15} (1977) 1966.
  doi:10.1103/PhysRevD.15.1966

\bibitem{Glashow:1976nt}
  S.~L.~Glashow and S.~Weinberg,
  Phys.\ Rev.\ D {\bf 15} (1977) 1958.
  doi:10.1103/PhysRevD.15.1958

\bibitem{Ecker:1987qp}
  G.~Ecker, W.~Grimus and H.~Neufeld,
  J.\ Phys.\ A {\bf 20} (1987) L807.
  doi:10.1088/0305-4470/20/12/010

\bibitem{Feldmann:2015nia}
  T.~Feldmann, T.~Mannel and S.~Schwertfeger,
  JHEP {\bf 1510} (2015) 007
  doi:10.1007/JHEP10(2015)007
  [arXiv:1507.00328 [hep-ph]].

\bibitem{Botella:1994cs}
  F.~J.~Botella and J.~P.~Silva,
  Phys.\ Rev.\ D {\bf 51} (1995) 3870
  doi:10.1103/PhysRevD.51.3870
  [hep-ph/9411288].

\bibitem{Davidson:2005cw}
  S.~Davidson and H.~E.~Haber,
  Phys.\ Rev.\ D {\bf 72} (2005) 035004
   Erratum: [Phys.\ Rev.\ D {\bf 72} (2005) 099902]
  doi:10.1103/PhysRevD.72.099902, 10.1103/PhysRevD.72.035004
  [hep-ph/0504050].

\bibitem{Jenkins:2009dy}
  E.~E.~Jenkins and A.~V.~Manohar,
  JHEP {\bf 0910} (2009) 094
  doi:10.1088/1126-6708/2009/10/094
  [arXiv:0907.4763 [hep-ph]].

\bibitem{Herren:2017}
F.~Herren, Master Thesis 2017,
Karlsruhe Institute of Technology

\bibitem{progdata}
\verb|https://www.ttp.kit.edu/preprints/2017/ttp17-046/|.

\bibitem{Christensen:2008py}
  N.~D.~Christensen and C.~Duhr,
  Comput.\ Phys.\ Commun.\  {\bf 180} (2009) 1614
  [arXiv:0806.4194 [hep-ph]].

\bibitem{Hahn:2000kx}
  T.~Hahn,
  Comput.\ Phys.\ Commun.\  {\bf 140} (2001) 418
  [hep-ph/0012260].

\bibitem{diss_salomon} 
J.~E.~Salomon, Dissertation 2012,
Karlsruhe Institute of Technology,\\
\verb|https://inspirehep.net/record/1351514/files/2294444.pdf|.
%

\bibitem{Nogueira:1991ex}
  P.~Nogueira,
  J.\ Comput.\ Phys.\  {\bf 105} (1993) 279.

\bibitem{Harlander:1997zb}
  R.~Harlander, T.~Seidensticker and M.~Steinhauser,
  Phys.\ Lett.\ B {\bf 426} (1998) 125
  [hep-ph/9712228].

\bibitem{Seidensticker:1999bb}
  T.~Seidensticker,
  hep-ph/9905298.

\bibitem{q2eexp}
  \verb|http://sfb-tr9.ttp.kit.edu/software/html/q2eexp.html|.

\bibitem{Kuipers:2012rf}
  J.~Kuipers, T.~Ueda, J.~A.~M.~Vermaseren and J.~Vollinga,
  Comput.\ Phys.\ Commun.\  {\bf 184} (2013) 1453
  doi:10.1016/j.cpc.2012.12.028
  [arXiv:1203.6543 [cs.SC]].

\bibitem{Larin:1991fz}
  S.~A.~Larin, F.~V.~Tkachov and J.~A.~M.~Vermaseren,
  preprint NIKHEF-H-91-18 (1991).

\bibitem{Steinhauser:2000ry}
  M.~Steinhauser,
  Comput.\ Phys.\ Commun.\  {\bf 134} (2001) 335
  [arXiv:hep-ph/0009029].

\bibitem{Steinhauser:1998cm}
  M.~Steinhauser,
  Phys.\ Rev.\ D {\bf 59} (1999) 054005
  doi:10.1103/PhysRevD.59.054005
  [hep-ph/9809507].

\bibitem{Gambino:1998ec}
  P.~Gambino, P.~A.~Grassi and F.~Madricardo,
  Phys.\ Lett.\ B {\bf 454} (1999) 98
  doi:10.1016/S0370-2693(99)00321-4
  [hep-ph/9811470].

\bibitem{Balzereit:1998id}
  C.~Balzereit, T.~Mannel and B.~Plumper,
  Eur.\ Phys.\ J.\ C {\bf 9} (1999) 197
  doi:10.1007/s100520050524
  [hep-ph/9810350].

\bibitem{Denner:2004bm}
  A.~Denner, E.~Kraus and M.~Roth,
  Phys.\ Rev.\ D {\bf 70} (2004) 033002
  doi:10.1103/PhysRevD.70.033002
  [hep-ph/0402130].

\bibitem{Santamaria:1993ah}
  A.~Santamaria,
  Phys.\ Lett.\ B {\bf 305} (1993) 90

\bibitem{Chowdhury:2015yja}
  D.~Chowdhury and O.~Eberhardt,
  JHEP {\bf 1511} (2015) 052
  doi:10.1007/JHEP11(2015)052
  [arXiv:1503.08216 [hep-ph]].

\end{thebibliography}
\end{document}